\documentclass[12pt]{article}
\usepackage[centertags]{amsmath}
\usepackage{graphicx}
\usepackage{upgreek}
\usepackage[makeroom]{cancel}
\usepackage{booktabs}
\usepackage{multicol}
\usepackage{multirow}
\usepackage{tabularx}
\usepackage{color}
\usepackage{fullpage}
\usepackage{subcaption}

\begin{document}

\title{Local and Nonlocal Shot Noise in High-$T_C$ Superconducting Nanowires}
\author{Corey Ostrove and L.  E. Reichl \\
Center for Complex Quantum Systems and Department of Physics\\
The University of Texas at Austin, Austin, Texas 78712\\}

\date{\today}

\maketitle\

\begin{abstract}

We obtain exact expressions for the local current and shot noise and the nonlocal shot noise in the first propagating channel of an NSN (normal-superconducting-normal) nanowire. Using high-$T_c$ cuprate superconductors (HTS) as our model systems we utilize a scattering theory approach to derive the scattering matrix for the systems in the high transparency limit and without the Andreev approximation. The local current and shot noise spectrum is calculated and plotted. In view of recent work in the area of cooper-pair splitting (CPS) devices, we investigate the nonlocal shot noise and relate the behavior of the noise as we change parameters such as the system size, the orientation of the order parameter and the biasing of the junction in the HTS NSN system. 

\end{abstract}

 \vspace{0.2cm}

\section{Introduction}

The possibility to create and control  entangled electron pairs in superconducting devices forms a basis for quantum information processors in solid state systems \cite{Celis,Veldhorst,Burset,Schindele,Hofstetter}. In superconductors, Cooper pairs can be broken apart to form entangled electron pairs that exit the device from separate leads. 
With the development of superconducting quantum information processors, and the development of high $T_c$ superconducting materials, there is a need to revisit the dynamical properties of superconducting heterostructures, particularly those formed from high-$T_c$ superconducting materials. They can be challenging to fabricate due to the difficulty of growing high quality single crystals at the nanoscale, and dimension limited superconductivity presents its own challenges in the form of dissipative finite size effects \cite{langer1967intrinsic,mccumber1970time}. 

High temperature cuprate superconductors (HTS) were discovered in 1986 \cite{bednorz1986possible} and, in the decades since, a substantial amount of experimental and theoretical development has occurred regarding the electronic transport properties of the materials \cite{tanaka1995theory,tanaka1996theory,kashiwaya2000tunnelling,tsuei2000pairing}.  Extensive experimental work has been performed on studying  mesoscopic d-wave superconductor structures (a review of which can be found in \cite{tafuri2005weak}). Additional experimental studies have probed current fluctuations and in particular the relation between shot noise and junction bias in {\it mesoscopic} d-wave structures \cite{constantinian2001experimental,constantinian2003shot,ovsyannikov2012josephson}. These shot noise studies find strong agreement with the theoretical predictions derived in \cite{anantram1996current} which were based the scattering matrix approach. Additional theoretical work on shot noise in d-wave superconductors can be found in \cite{cuevas2002shot} and a general review of shot noise in mesoscopic systems can be found in \cite{blanter2000shot}.

In recent years, advances in fabrication techniques and HTS material science have enabled the manufacture of high quality single crystal nanowires. These nanowires allow for the study of dimensionality-limited superconductivity  with the smallest of the nanowires operating in the quasi 1-d limit. Applications have been proposed for using superconducting nanowire devices to improve the quantum efficiency of single photon detectors \cite{bartolf2010current} and for improving the flux sensitivity of SQUIDs due to their high critical magnetic field strengths and small coherence lengths \cite{arpaia2014ultra,schwarz2015low}

 Of all of the HTS, the most effort has been applied to the manufacturing and characterization of nanoscale YBCO  ($YBa_2 Cu_3 O_{7-x}$). The state-of-the-art for YBCO nanowire size is ~10nm \cite{xu2008long}. Nanowires of this size are on order of the coherence length for YBCO, ${\xi}=3.4$ nm  \cite{wesche2015physical},  and smaller than the magnetic penetration depth of ${\lambda}=26 nm$ \cite{wesche2015physical}. In addition to YBCO, nanowires with widths as small as 80nm have been constructed from LSCO ($La_{2-x}Sr_xCuO_4$) \cite{litombe20142} as well as nanowires with widths between 400-600nm for BSCCO ($Bi_2(Sr,La)_2CuO_y$)  \cite{duarte2013electrospinning}. These dimensions are still at least an order of magnitude larger than the coherence length, but it is conceivable that it will be possible within the next few years to reliably construct nanowires with sizes on the order of the coherence length for a number of cuprate HTS. 

The usefulness of superconductors as an entanglement source is limited by the ability to spatially separate the entangled electrons comprising the Cooper pairs, and thus is determined by the spatially nonlocal transport properties of the structure which couple different normal metal leads in the system \cite{lesovik}. The primary mechanism responsible for the creation of entangled electron pairs in NSN structures is the crossed Andreev reflection (CAR) process in which a particle or hole  impinges on the superconducting region in one lead and is transmitted as a hole or particle, respectively,  in the other lead \cite{lesovik,bouchiat}. An important measure for entanglement generation in these structures, Cooper pair splitting efficiency $\eta$, can be related to the strength of this process in relation to the other non-entanglement generating processes such as specular reflection and transmission of particles or of holes, without a change in particle type \cite{Burset}. Alternatively, it is possible to relate the efficiency of formation of entangled pairs  to the nonlocal shot noise spectrum of the structure, which is one of the focuses in this paper \cite{Celis,buttikersign}. As a result of the interest in Cooper pair splitter design a large amount of work has been done theoretically (see for example, \cite{Veldhorst,Burset,Yeyati,Freyn}) and experimentally (see \cite{Schindele,Hermann,Hosfstetterphysrev}) to characterize the nonlocal properties of various NSN structures. Most of this work so far has focused on standard superconductors. Some recent work by Celis Gil et. al. \cite{Celis} has worked out some theory for systems with anistropic order parameters, such as HTS.

In this paper we adopt the scattering theory framework for quantum electronic transport in nanowires, developed by Landauer \cite{Landauer} and Buttiker \cite{Buttiker}. However, we  generalize it to include inelastic scattering processes due to the superconducting segment of the nanowire by adapting scattering theory developed by  Demers and Griffin and by Blonder, Tinkham, and Klapwijk  \cite{Demers,BTK} to nanometer scale quantum wires.  We obtain exact expressions without the Andreev approximation \cite{Andreev} for current and both local and nonlocal shot noise and use these expressions to get numerical results for a HTS nanowire NSN heterostructure using parameters for LSCO. In Sect. \ref{sec:models}, we derive the field operators for electrons in the left lead, and include the effect of scattering from the superconducting part of the nanowire.  In Sect.  \ref{sec:supercond}, we review the Bogoliubov-de Gennes theory of the superconductor as it applies to the nanowire. In Sect. \ref{sec:scattamp}, we describe the scattering theory used to obtain exact expressions (given in Appendix B)  for the scattering amplitudes.  In Sect. \ref{sec:numbers}, we list materials that have a coherence length of order of the width of the nanowire and which could propagate a superconducting current through the nanowire.   In Sect. \ref{sec:shot noise}, we give exact expressions for the average current in the wire, and for the local shot noise in the wire along with numerical results for LSCO. In Sect. \ref{sec:nonlocal shot noise} we get the exact expression for the nonlocal noise and present numerical results for the noise in various parameter regimes for LSCO. In Sect. \ref{sec:conclude}, we make some concluding remarks.

\section{Scattering Theory Methods}
\label{sec:scattering_theory}

Our analysis of the HTS NSN junction is based on the scattering theory approach. In this section we develop the theory which we use in order to calculate the scattering matrix for the system. The scattering matrix is found in the high transparency limit and without use of the Andreev approximation. In Appendix B we give the analytical expressions for the scattering matrix elements as a function of the system parameters.

\subsection{Electron Dynamics in the Left Lead}
\label{sec:models}

We consider an NSN (normal-superconducting-normal) composite nanowire. The left lead is normal with length $L_L$ and is connected to a very low temperature electron reservoir with Fermi energy $E^L_F$. The right lead is normal, with length $L_R$ and is connected to a different low temperature electron reservoir with Fermi energy $E^R_F$. 
The central part of the nanowire is superconducting with length $L_S$. The current flows along the z-axis. The superconducting part begins at $z=0$ and ends at $z=L_S$. 
 For simplicity, we assume that the nanowire  is a rectangular tube with infinitely hard walls. at $x=0$, $x=L_x$, $y=0$, $y=L_y$, where $L_x>L_y$.  We also assume that the nanowire is much longer than it is wide.  

We first consider the electron current in the left lead. The field operator for  electrons in the left lead can be written 
\begin{eqnarray}
{\hat {\psi}}(x,y,z,t)=~~\frac{1}{\sqrt{{\rm L}_L}}{\sum_{n_z=1}^{\infty}} {\sum_{{\nu}}}{\rm e}^{-iE_{n_z,\nu}t/{\hbar}}~{\phi}_{\nu}(x,y)  \left( {\hat a}^L_{q_{n},\nu}{\rm e}^{iq_{n}x}+{\hat b}^L_{q_{n},\nu}{\rm e}^{-iq_{n}x} \right),~~~~~~~~~~~~~
\label{field0}
\end{eqnarray}
where $\nu={\{}n_x,n_y{\}}$ denotes the transverse modes, ${\rm L}_L$ is the length of the left lead along the z-axis, and  
\begin{eqnarray}
{\phi}_{\nu}(x,y)=\frac{2}{\sqrt{{\rm L}_x{\rm L}_y}}{\sin} \left( \frac{n_x{\pi}x}{{\rm L}_x}\right) {\sin} \left( \frac{n_y{\pi}y}{{\rm L}_y}\right)
\end{eqnarray}
 is the normalized wave function for the transverse modes  We assume that $L_L \gg L_x>L_y$.  The operator $ {\hat a}^L_{q_{n},\nu}$  annihilates  electrons that are emitted from the left reservoir and travel to the right in the ${\nu}$\textsuperscript{th} transverse channel, and $ {\hat b}^L_{q_{n},\nu}$ annihilates   electrons that travel to the left, either due to reflection or because they are emitted from the electron reservoir on the right.  We assume that no impurities are present that can couple different transverse channels.  

  The total energy of the electrons in the ${\alpha}\textsuperscript{th}$ normal lead $({\alpha}=L,R)$  can be written $E^{\alpha}_{n,n_x,n_y}=e^{{\alpha}}_{n}+E^{{\alpha}}_{n_x,n_y}$, where $e^{{\alpha}}_{n}=\frac{{\hbar}^2}{2m_{{\alpha}}}(q^{{\alpha}}_{n})^2$,     $q^{{\alpha}}_{n}=\frac{2{\pi}n}{{\rm L}_{\alpha}}$ ($n=0,...,{\infty}$) is the wave vector associated with electron motion along the z-axis (longitudinal motion) in the ${\alpha}$\textsuperscript{th}  lead  and $E^{{\alpha}}_{n_x,n_y}$ is the energy associated with the transverse degrees of freedom in the ${\alpha}$\textsuperscript{th} lead.  For simplicity, we specialize to the case where only the first transverse modes, $n_z=1,n_y=1$,  are excited, and we adjust the Fermi energy so it lies approximately  midway between the highest energy $E^{{\alpha}}_{2,1}$ and the lowest energy $E^{{\alpha}}_{1,1}$  of the first propagating channel.  Thus, the electron reservoirs on the left and right  sides  of the waveguide have Fermi energies $E^{\alpha}_F{\approx}(E^{{\alpha}}_{1,1}+E^{{\alpha}}_{2,1})/2$. The Fermi energies of the reservoirs, on the left and right, might be slightly different, but we assume the difference is small.
   
For the case when only the first propagating channel in the left lead is  energetically allowed,  $q^{L}_{n}=\frac{2{\pi}n}{{\rm L}_L}$ is the wave vector for longitudinal mode $n$, and  $0{\leq}\frac{{\hbar}^2}{2m_{L}}(q^{L}_{n})^2{\leq}(E_{2,1}^{\ell,tr}-E_{1,1}^{\ell,tr})$, so  $0{\leq}n{\leq}n_{max}$,  where $\frac{{\hbar}^2}{2m_{L}}(q^{L}_{n_{max}})^2 = (E_{2,1}^{L,tr}-E_{1,1}^{L,tr})$.
  For $n=0$ the total energy $E^{L}_{0}=E^{L,tr}_{1,1}$.  For $n=n_{max}$ the total energy $E_{n_{max}}=E^{L,tr}_{2,1}$.
For $n=n^{L}_{F}$ the total energy $E_{n^{L}_{F}}=E^{L}_F=\frac{{\hbar}^2}{2m}(q^{L}_{n_F})^2+E^{L,tr}_{1,1}$.   It is useful to define ${\epsilon}^{L}_F=\frac{{\hbar}^2}{2m_{L}}(q^{L}_{n_F})^2=E^{L}_F-E^{L,tr}_{1,1}=(E_{2,1}^{L,tr}-E_{1,1}^{L,tr})/2$.

We next divide particle  states in the left lead according to whether their energy is greater or less than ${\epsilon}^{L}_F$.  We define ``B" states  with energy below the Fermi energy $e_{B,n_{L}}={\epsilon}^{L}_F-\frac{{\hbar}^2}{2m_{L}}q_{B,n_{L}}^2$ with $0{\leq}n_{L}{\leq}n^{L}_F$.  
We define ``A" states with energy above the Fermi energy $e_{A,n_{L}}=\frac{{\hbar}^2}{2m_{L}}q_{A,n_{L}}^2-{\epsilon}^{L}_F$ with $n^{L}_F{\leq}n_{L}{\leq}n^{L}_{mx}$. 
The longitudinal wave vectors in the {\it left} lead can be written
\begin{equation}
\begin{aligned}[-1]
q_{A,n_{L}} & =\sqrt{\frac{2m_{L}}{{\hbar}^2}\left( e_{A,n_{L}}+{\epsilon}^{L}_F\right)}, & & n^{L}_F {\geq} n_{L} {\geq} n^{L}_{max}\\
q_{B,n_{L}} & =\sqrt{\frac{2m_{L}}{{\hbar}^2}\left( {\epsilon}^{L}_F-e_{B,n_{L}}\right)}, & &  0 {\leq} n_{L} {\leq} n^{L}_F
\end{aligned}
\end{equation}

We next change the summation to an integration over energy, since that is common to all terms \cite{Reichl}. 
First note  that
\begin{eqnarray}
{\sum_{n=0}^{n_{max}}}{\rightarrow}{\int_{0}^{n_{max}}}dn{\rightarrow}{\int_{0}^{2{\epsilon}_F}}de\frac{dn}{de} =\frac{{\rm L}_L}{2{\pi}}\frac{m_{L}}{{\hbar}^2}{\int_{0}^{2{\epsilon}_F}}de\frac{1}{q_e^{L}},~~
~~~\end{eqnarray}
where $q_e^{L}=\sqrt{2m_{L}e}/{\hbar}$, and $2{\epsilon}_f=(E^{tr}_{2,1}-E^{tr}_{1,1})$, which is the energy interval of the first channel.
Also note that
the creation and annihilation operators for particles can be written
\begin{eqnarray}
{\sum_{n=0}^{n_{max}}}{\hat a}^{\dagger}_{q_{n}}{\hat a}_{q_{n}}=\frac{{\rm L}_L}{2{\pi}}\frac{m_L}{{\hbar}^2}{\int_{0}^{2{\epsilon}_F}}de\frac{1}{q_e}~{\hat a}^{\dagger}_{q_{n}}{\hat a}_{q_{n}}
{\equiv}{\int_{0}^{2{\epsilon}_F}} de~ {\hat a}^{{\dagger}}_{e}{\hat a}_{e},
\end{eqnarray}
where ${\hat a}^{\alpha}_{e}=\sqrt{ \frac{{\rm L}_L}{2{\pi}}\frac{m_{L}}{{\hbar}^2q_e}}~{\hat a}_{q_n}$. 
The longitudinal wave vectors in the {\it left} lead can now be written
\begin{equation}
\begin{aligned}
q^{L}_{A}&=\sqrt{\frac{2m_{L}}{{\hbar}^2}\left( e+{\epsilon}_F\right)} ,~~~0{\leq}e{\leq}{\epsilon}_F\\
~~q^L_{B}&=\sqrt{\frac{2m_{L}}{{\hbar}^2}\left( {\epsilon}_F-e\right)},~~~0{\leq}e{\leq}{\epsilon}_F,
\nonumber\\
\end{aligned}
\end{equation}

We next rewrite the field operators in terms of continuous particle  energies . Notice that the wave vectors of type A and B particles have different dependence on energy, $e$. For an NNN nanowire, they are uncoupled but, for the NSN nanowire they are coupled due to the inelastic Andreev scattering process through the superconducting segment of the wire \cite{Andreev}. We  first write the field operator for the A particle states in the first propagating channel. Eq. (\ref{field0}), takes the form
\begin{eqnarray}
{\hat {\psi}}^{L}_A(x,t)
=\sqrt{\frac{1}{2{\pi}}\frac{m_{L}}{{\hbar}^2}}{\int_{0}^{{\epsilon}_F}}de \frac{1}{\sqrt{q_{A}^{L}}}~  {\rm e}^{-i({\epsilon}_F+e)t /{\hbar} }
{\big(}  ~{\hat a}^L_{A,e}{\rm e}^{iq_A^{L}z}+ ~{\hat b}^L_{A,e}{\rm e}^{-iq_A^{L}z}{\big)}~~~~~~~~~~~~~~~
\label{field5}
\end{eqnarray}
Next write the part of the field operator for the B particle states. Eq. (\ref{field0}), takes the form
\begin{eqnarray}
{\hat {\psi}}^L_B(x,t)
=\sqrt{\frac{1}{2{\pi}}\frac{m_{L}}{{\hbar}^2}}{\int_{0}^{{\epsilon}_F}}de\frac{1}{\sqrt{q_{B}^{L}}}~  {\rm e}^{-i({\epsilon}_F-e)t /{\hbar} }
{\big(} ~{\hat a}^L_{B,e}{\rm e}^{-iq_{B}^Lz}+~{\hat b}^L_{B,e}{\rm e}^{iq_B^Lz}{\big)}~~~~~~~~~~~~~~~~~~
\label{field6}
\end{eqnarray}
Note that similar expressions for the particle field operators can be derived for the normal right lead.
For these two field operators,  the wave vectors $q_A^{L}$ and $q_B^{L}$ depend on energy $e$ differently, and the character of the A and B type excitations is different, as can be seen from the behavior of  thermal averages.  The thermal distributions for type A and B excitations are 
$f_A({e})=\frac{1}{(1+{\rm e}^{{\beta}{ e}})}$  and $ f_B({ e})=\frac{1}{(1+{\rm e}^{-{\beta}{ e}})}$, respectively (see Appendix A). They describe the distribution of particles for energy moving away from the Fermi energy. 

 We now let ${\hat a}^{L \dagger}_{A}{\rightarrow}{\hat a}^{L \dagger}_{p}$ and let 
${\hat a}^{L }_{B}{\rightarrow}{\hat a}^{L \dagger}_{h}$, where ${\hat a}^{L \dagger}_{h}$ creates a hole at energies below the Fermi energy. 
The thermal averages are then $N_p(e)={\langle}{\hat a}^{L \dagger}_{p}{\hat a}^{L}_{p} {\rangle}=(1+{\rm e}^{{\beta}{ e}})^{-1}$ and $N_h(e)={\langle}{\hat a}^{L \dagger}_{h}{\hat a}^{L}_{h} {\rangle}=(1+{\rm e}^{{\beta}{ e}})^{-1}$.
The field operator, Eq. (\ref{field0}), can be written in terms of particle and hole excitations, and we explicitly include the scattering processes that give rise to left-traveling electron in the left lead. The field operator then takes the form
\begin{equation}
\begin{aligned}
{\hat {\psi}}_L(z,t)
& =\sqrt{\frac{1}{2{\pi}}\frac{m_{L}}{{\hbar}^2}}   {\rm e}^{-i{\epsilon}_Ft /{\hbar} }{\int_{0}^{{\epsilon}_F}}de {\bigg[}\frac{1}{\sqrt{q_{p}^{L}}}~  {\rm e}^{-iet /{\hbar} }
{\bigg(}  ~{\hat a}^L_{e,p}{\rm e}^{iq_p^{L}z}+
~r^{LL}_{pp}(e){\hat a}^L_{e,p}{\rm e}^{-iq_p^{L}z}\\
&+t^{LR}_{pp}(e) {\hat a}^R_{e,p}{\rm e}^{-iq_p^{L}z}
+ r^{LL}_{ph}(e) {\hat a}^L_{e,h}{\rm e}^{-iq_p^{L}z}+ t^{LR}_{ph}(e){\hat a}^R_{e,h}{\rm e}^{-iq_p^{L}z} {\bigg)}\\
& +\frac{1}{\sqrt{q_{h}^{L}}}~  {\rm e}^{+iet /{\hbar} }
{\bigg(} ~{\hat a}^{L\dagger}_{e,h}{\rm e}^{-iq_h^{L}z}+  r^{LL}_{hp}(e){\hat a}^{L\dagger}_{e,p}{\rm e}^{iq_h^{L}z} \\
&+t^{LR}_{hp}(e) {\hat a}^{R \dagger}_{e,p} {\rm e}^{iq_h^{L}z} 
 +  r^{LL}_{hh}(e) {\hat a}^{L \dagger}_{e,h}{\rm e}^{iq_h^{L}z} +  t^{LR}_{hh}(e){\hat a}^{R \dagger}_{e,h}{\rm e}^{iq_h^{L}z}   {\bigg)}~{\bigg]}
 \end{aligned}
 \label{field1}
\end{equation}
 The operator $ {\hat a}^L_{{e},p} ~( {\hat a}^L_{{ e},h} )$ destroys a particle (hole) coming from the left reservoir. The operator $ {\hat a}^R_{{ e},p}~( {\hat a}^R_{{ e},h})$ destroys a particle (hole) coming from the right reservoir.  The scattering  amplitudes $r^{LL}_{pp}$, $r^{LL}_{ph}$, $r^{LL}_{hp}$, and $r^{LL}_{hh}$, are reflection probability amplitudes for $p{\rightarrow}p$,  $h{\rightarrow}p$, $p{\rightarrow}h$, and $h{\rightarrow}h$, respectively.  The scattering  amplitudes $t^{LR}_{pp}$, $r^{LR}_{ph}$, $t^{LR}_{hp}$, and $t^{LR}_{hh}$, are transmission probability amplitudes for $p{\rightarrow}p$,  $h{\rightarrow}p$,, $p{\rightarrow}h$, and $h{\rightarrow}h$, respectively, from the right lead to the left lead (these scattering amplitudes are defined in Appendix B).

\subsection{Superconductor Dynamics}
\label{sec:supercond}

The dynamics of the superconducting part of the waveguide are determined using Bogoliubov-de Gennes (BdG) theory \cite{deGennes}.  We assume that  the superconducting nanowire has the same cross section as the normal parts of the wire, and that the superconducting lead has length ${\rm L}_s$ with $L_s{\gg}L_x$ and $L_s{\gg}L_y$. We also assume that the temperature of the system is sufficiently low, so as to make the effects of phase slipping negligible \cite{langer1967intrinsic,mccumber1970time}. Energy eigenstates in the superconductor are solutions of the Bogoliubov-de Gennes equations.  

\begin{equation}
\left(\begin{array}{cc}-\frac{{\hbar}^2{\nabla}_{\bf r}^2}{2m_s}-E^s_F & {\Delta}\\
{\Delta}&\frac{{\hbar}^2{\nabla}_{\bf r}^2}{2m_s}+E^s_F \\
\end{array}\right)\left(\begin{array}{c}u({\bf r})\\ v({\bf r}) \\
\end{array}\right)={ e}\left(\begin{array}{c}u({\bf r})\\ v({\bf r}) \\
\end{array}\right).
\label{BdG1}
\end{equation}
where $m_s$ is the electron effective mass in the superconductor and  ${e}$ denotes the excitation energies, measured relative to the Fermi energy.  The components of the energy eigenfunction, $u({\bf r})$ and  $v({\bf r})$, respectively describe particle and hole behavior in the superconductor. BdG theory explicitly couples particles and holes via the gap function ${\Delta}$ which is the binding energy of the particle-hole pairs that form the condensate and is the order parameter of the superconductor. For the most common HTS materials, such as the cuprates, the order parameter is actually anisotropic, with the most common form for the anisotropy having what it known as d-wave pairing symmetry. The functional form of the d-wave order parameter is given by $\Delta=\Delta(\theta)=\Delta_0 \cos{2(\theta)}$, where $\Delta_0$ is the maximum value of the gap and $\theta$ is the direction of travel through the superconductor as measured with respect to the $a$-crystal axis. In this paper we'll often refer to $\theta$ as the ``orientation'' of the order parameter and the angular dependence of the quantities which depend on $\Delta$ is implied.

We now write the energy eigenfunction for the nanowire
\begin{equation}
\left(\begin{array}{c}u({\bf r})\\ v({\bf r}) \\
\end{array}\right)={\sum_{n_x=1}^{\infty}} {\sum_{n_y=1}^{\infty}} {\sum_{n_z=-{\infty}}^{\infty}} \left(\begin{array}{c}u_{\bf n}\\ v_{\bf n} \\
\end{array}\right)  {\rm exp}\left(ik_{z}z \right) {\phi}_{n_x,n_y}(x,y)\end{equation}
where $k_{n_s}{\equiv}\frac{2{\pi}n_s}{L_s}$ and ${\bf n}=(n_s,n_y,n_z)$.  Plugging into Eq. (\ref{BdG1}), we get
\begin{equation}
\left(\begin{array}{cc}\frac{{\hbar}^2k_{n_s}^2}{2m_s}+E_{n_x,n_y}^{s,tr}-E^s_F & {\Delta}\\
{\Delta}&-\frac{{\hbar}^2k_{n_s}^2}{2m_s}-E_{n_x,n_y}^{s,tr}+E^s_F \\
\end{array}\right)   \left(\begin{array}{c}u_{\bf n}({ e})\\ v_{\bf n}({ e}) \\
\end{array}\right)={ e}\left(\begin{array}{c}u_{\bf n}({ e})\\ v_{\bf n}({ e}) \\
\end{array}\right).
\label{BdG2}
\end{equation}
where $E_{n_x,n_y}^{s,tr}=\frac{{\hbar}^2}{2m_s} \left( \frac{n_x^2{\pi}^2}{L_x^2}+\frac{n_y^2{\pi}^2}{L_y^2}\right)$ denotes the energy in the transverse modes. 
We will  again specialize to the case where only the first transverse modes are excited.  Then $n_x=n_y=1$, and  the Fermi energy $E^s_F$ is confined to the values $E_{1,1}^{\ell,tr}<E^R_F< E^{\ell,tr}_{2,1}$ to allow propagation of electrons in the first channel   of the {\it normal} conducting wire.  
 
The BdG equation for the energy ${e}$ of the excitations (as measured relative to the Fermi energy)  can now be written
\begin{equation}
\left(\begin{array}{cc} \frac{{\hbar}^2k^2}{2m_s}+E_{1,1}^{,tr}-E^R_F& {\Delta}\\
{\Delta}&-\frac{{\hbar}^2k^2}{2m_s}-E_{1,1}^{tr}+E^R_F \\
\end{array}\right) \left(\begin{array}{c}u({e})\\ v({ e}) \\
\end{array}\right)={ e}\left(\begin{array}{c}u({ e})\\ v({e}) \\
\end{array}\right).
\label{BdG3}
\end{equation}
To simplify notation, set ${\epsilon}_F=E^R_F-E_{1,1}^{s,tr}$, as was done in the normal part of the nanowire. Then the BdG equation can be written 
\begin{equation}
\left(\begin{array}{cc}\frac{{\hbar}^2k^2}{2m_s}-{\epsilon}_F& {\Delta}\\
{\Delta}&-\frac{{\hbar}^2k^2}{2m_s}+{\epsilon}_F \\
\end{array}\right) \left(\begin{array}{c}u({ e})\\ v({ e}) \\
\end{array}\right)={ e}\left(\begin{array}{c}u({ e}) \\ v({ e}) \\
\end{array}\right).
\label{BdG4}
\end{equation}
%
%
The energy eigenvalues for excitations are $e=\sqrt{K^2+{\Delta}^2}$.  

The particle-like excitations have wave vector and energy eigenvector
\begin{eqnarray}
k_p=\sqrt{\frac{2m_s}{{\hbar}^2}}\sqrt{{\epsilon}_F +\sqrt{{ e}^2-{\Delta}^2}}~~{\rm and}~~ \left(\begin{array}{c}u_p({ e})\\ v_p({ e}) \\
\end{array}\right) =\left(\begin{array}{c}u_o\\ v_o \\
\end{array}\right) 
\end{eqnarray}
respectively, where
\begin{eqnarray}
u_{o}^2=\frac{1}{2}\left(1+\frac{\sqrt{{ e}^2-{\Delta}^2}}{{ e}}\right)~~~{\rm and}~~~
v_{o}^2=\frac{1}{2}\left(1-\frac{\sqrt{{ e}^2-{\Delta}^2}}{{ e}}\right).
\label{pex}
\end{eqnarray}
The hole-like excitations have wave vector and energy eigenvector
\begin{eqnarray}
k_h=\sqrt{\frac{2m_s}{{\hbar}^2}}\sqrt{{\epsilon}_F -\sqrt{{ e}^2-{\Delta}^2}}~~{\rm and}~~ \left(\begin{array}{c}u_h({ e})\\ v_h({ e}) \\
\end{array}\right) =\left(\begin{array}{c}v_o\\ u_o \\
\end{array}\right) 
\end{eqnarray}
respectively.

\subsection{The Scattering System}
\label{sec:scattamp}

In this and subsequent sections, we express all quantities in atomic units. 
For energy, the atomic unit is, $1.0 E_h=27.216~eV=4.36{\times}10^{-18}~J$and is measured in hartrees. For length, the atomic unit is $a_B= 1.0~ bohr =0.052917~nm$, where $a_B$ is the Bohr radius, and for velocity the atomic unit is $v_B=2.1877{\times}10^6~m/s$. In atomic units, the electron mass is $m_e=1$ and Planck's constant is ${\hbar}=1$. 

Let us assume that both a particle and a hole are incident from the thermal reservoirs on the left and on the right.  
The states in the left normal region, ${\psi}_L(z)$, the middle superconductor region,  ${\psi}_S(z)$, and the right normal region, ${\psi}_R(z)$, are respectively

\begin{equation}
\begin{aligned}
{\psi}_L(z)& =\frac{A_L^p}{\sqrt{q_p^{L}}} \left[ \begin{array}{c} 1\\0 \end{array} \right] {\rm e}^{iq_p^{L}z}+\frac{B_L^h}{\sqrt{q_h^{L}}} \left[ \begin{array}{c} 0\\1 \end{array} \right] {\rm e}^{-iq_h^{L}z}\\
&\hspace{8.4em}+\frac{C_L^p}{\sqrt{q_p^{L}}} \left[ \begin{array}{c} 1\\0 \end{array} \right] {\rm e}^{-iq_p^{L}z}+\frac{D_L^h }{\sqrt{q_h^{L}}}\left[ \begin{array}{c} 0\\1 \end{array} \right] 
{\rm e}^{+iq_h^{L}z}, \\
{\psi}_S(z)&=\frac{A_S^p}{\sqrt{k_p}} \left[ \begin{array}{c} u_o\\ v_o \end{array} \right] {\rm e}^{ik_pz}+\frac{B_S^h}{\sqrt{k_h}} \left[ \begin{array}{c} v_o\\u_o \end{array} \right] {\rm e}^{-ik_hz}\\
&\hspace{8.6em}+\frac{C_S^p}{\sqrt{k_p}} \left[ \begin{array}{c} u_o\\v_o \end{array} \right] {\rm e}^{-ik_pz}+\frac{D_S^h}{\sqrt{k_h}} \left[ \begin{array}{c} v_o\\u_o\end{array} \right] 
{\rm e}^{+ik_hz}, \\
{\psi}_R(z)&=\frac{A_R^p}{\sqrt{q^R_p}} \left[ \begin{array}{c} 1\\0 \end{array} \right] {\rm e}^{iq^R_pz}+\frac{B_R^h}{\sqrt{q^R_h}} \left[ \begin{array}{c} 0\\1 \end{array} \right] {\rm e}^{-iq^R_hz}\\
&\hspace{8.5em}+\frac{C_R^p}{\sqrt{q^R_p}} \left[ \begin{array}{c} 1\\0 \end{array} \right] {\rm e}^{-iq^R_pz}+\frac{D_R^h}{\sqrt{q^R_h}} \left[ \begin{array}{c} 0\\1 \end{array} \right] 
{\rm e}^{+iq^R_hz}.
\end{aligned}
\label{scatsys}
\end{equation}

\noindent where (in atomic units) $q^{\alpha}_{\beta} =\sqrt{2m_{\alpha}^*\left( { e} \pm {\epsilon}^{\alpha}_F\right)}$ with $\alpha=L,R$ and $\beta=p,h$ and where we use the upper sign for particles and the lower sign for holes. The choice of normalization in Eqs. (\ref{scatsys}) ensures that our system carries unit current \cite{beenaker}. Using the Eqs. (\ref{scatsys}), we can  determine the probability amplitudes for electrons to transmit through and be reflected from the superconducting part of the nanowire. 
We are working in the high transparency limit and so we require that the wave functions in Eqs. (\ref{scatsys}), be continuous at $z=0$ and $z={\rm L}_s$, and that the slopes be continuous at those points. Thus
\begin{equation}
{\psi}_L(0)={\psi}_S(0),~~~{\psi'}_L(0)={\psi'}_S(0),~~~{\psi}_S(L_S)={\psi}_R(L_S),~~~{\psi'}_S(L_S)={\psi'}_R(L_S)
\end{equation}
Using these equations, we can eliminate the coefficients $A_S^p,~ B_S^h,~C_S^p,~D_S^h$ and write $C_L^p, ~D_L^h,~A_R^{p},~B_R^{h}$ as functions of  $A_L^p, ~B_L^h,~C_R^{p},~D_R^{h}$.  This gives the matrix equation
\begin{equation}
\left(
\begin{array}{c}
C_L^p \\
D_L^h \\
A_R^{p'} \\
B_R^{h'} \\
\end{array}
\right)=\left(\begin{array}{cccc}
r^{LL}_ {pp} &r^{LL}_ {ph} &t^{LR}_ {pp'}&t^{LR}_ {ph'}\\
r^{LL}_ {hp}& r^{LL}_ {hh}&  t^{LR}_ {hp'}&t^{LR}_ {hh'} \\
t^{RL}_ {p'p}&t^{RL}_ {p'h}& r^{RR}_ {p'p'}& r^{RR}_ {p'h'}\\
t^{RL}_ {h'p}&t^{RL}_ {h'h}& r^{RR}_ {h'p'}& r^{RR}_ {h'h'} \end{array} \right){ \cdot}\left(
\begin{array}{c}
A_L^p \\
B_L^h \\
C_R^{p'} \\
D_R^{h'} \\
\end{array}
\right)
\label{S-matrix}
\end{equation}
Explicit expressions for these transmission and reflection amplitudes are given in Appendix B.

In Fig. \ref{fig:1}, we plot the magnitude of the transmission and reflection amplitudes for the LSCO nanowire ($La_{2-x}Sr_xCu_4$) using parameters from Table I, with a superconducting segment of lengths $L_S=5000~au$ (solid line) and $L_S=500~au$ (dashed line).  Fig. \ref{fig:1}.a shows the amplitude for a particle (hole) to leave the left reservoir and reflect back into the left lead as a particle (hole).  Fig. \ref{fig:1}.b  shows the  amplitude for a particle (hole) to leave the right reservoir and transmit through the superconductor and enter the left lead as a particle (hole). Fig. \ref{fig:1}.c shows the amplitude for a particle (hole) to leave the left reservoir and reflect back into the left lead as a hole (particle). Fig. \ref{fig:1}.d shows the amplitude for a particle (hole) to leave the right reservoir and transmit into the left lead as a hole (particle).  When the energy $e$ is less  than the gap energy, the probability of Andreev reflection and of particle or hole transmission through the superconductor is nearly unity.

\section{Numbers for Some High-$T_c$ materials}
\label{sec:numbers}

We now compute the current and shot noise for some high-$T_c$ superconductors and determine if any can transmit a current in the first propagating channel of our nanowire and what the size of the nanowire would need to be.  We express all quantities in atomic units. 

Let us consider some typical high $T_c$ materials.  The gap in cuprates is anisotropic, and can vary greatly depending on orientation  \cite{sigrist2005introduction, maki1998introduction}. As an approximation we use estimates for the maximum gap value $(\theta=0)$ based on BCS. The BCS estimate of the superconducting gap is ${\Delta}=1.77k_BT_c$, where $k_B$ is Boltzmann's constant. In atomic units this is 
\begin{equation}
{\Delta}=\left(5.572{\times}10^{-6}~\frac{E_h}{K} \right)T_c,
\end{equation}
where $T_c$ is measured in Kelvins. The Fermi velocity is 
\begin{equation}
v_F={\xi}{\pi}{\Delta}=\left(1.75{\times}10^{-5}~\frac{E_h}{K}\right){\xi}T_c,
\end{equation}
where the coherence length ${\xi}$  is measured in atomic units and $T_c$ in Kelvins. The Fermi energy is 
\begin{equation}
E_F=\frac{1}{2}m^*v_F^2=\left(1.532{\times}10^{-10}~\frac{E_h}{K^2} \right)~m^*{\xi}^2T_c^2,
\label{FermiEn}
\end{equation}
where the effective mass, $m^*$, and $\xi$ are measured in atomic units and $T_c$ in Kelvins.
The Fermi energy lies at the center of the energy interval defined by the first propagating channel inn the waveguide.  We will assume that $L_x=1.2L_y$, so the second transverse channel opens along the x-direction.   The value of $L_y$ can be determined from the condition that
\begin{equation}
E_F=\frac{1}{2}(E_{1,1}+E_{2,1})=\frac{1}{4m^*}\left(5\left(\frac{{\pi}}{1.2L_y}\right)^2+2\left(\frac{{\pi}}{L_y}\right)^2 \right).
\end{equation}
If we use Eq. \ref{FermiEn}, we obtain 
\begin{equation}
L_y=\frac{2.97{\times}10^5}{m^*{\xi}T_c}~au{\cdot}K.
\end{equation}
for the width $L_y$ of the nanowire.

We can now evaluate these quantities for several known high $T_c$ superconducting materials, where for simplicity we'll take the effective mass $m^*=1$ in all three materials.
The  Fermi distribution for particles and holes is $N_p(e)=N_h(e)=(1+{\rm e}^{e/k_BT})^{-1}$ (they describe the distribution of particles and hole, respectively in different parts of the energy axis). We can obtain an estimate for the temperature at which the current flow would be maximized. Let us require that 
$(1+{\rm e}^{{\delta}E/(k_BT_0)})^{-1}=0.001$, so the distribution of particles and holes has decayed to approximately zero a distance ${\delta}E$ from the Fermi energy (on their respective sides of the energy axis). This gives $T_0=45729{\delta}E$ in Kelvin.
For this temperature, the Fermi distributions $N_p(e)$ and $N_h(e)$ range in value from $0.5$ to 0.001 in the energy interval ${\delta}E$, which is the energy half-width of the first propagating channel. Values for all these parameters for a variety of high-$T_c$ materials are given in Table 1.

\begin{table}
	\begin{center}
		\resizebox{\textwidth}{!}{
			\begin{tabular}{cccc}
				\hline
				\hline
				Material &$YBa_2 Cu_3 O_{7-x} $(a) & $La_{1.45}Nd_{0.4}Sr_{0.15}CuO_{4+{\delta}}$(a)&$Nd_{1.85}Ce_{0.15}CuO_{4-{\delta}}$(a)\\
				\hline
				$T_c~(K)$ & 88.8 & 24 &26.2 \\
				$\xi~(a_B)$ & 64.3 & 91 &101 \\
				$\Delta~(E_H)$ &0.000495 & 0.000134 &0.000146 \\
				$E_F~(E_H)$ & 0.00499 & 0.000726 &0.00107 \\
				${\delta}E~(E_H)$ & 0.00190 & 0.000276 &0.000406 \\
				$L_y~(a_B)$ & 52 & 136.4 &113 \\
				$T_0~(K)$ & 43.4 & 6.32 &9.28 \\
				\hline
				\hline
				Material &$La_{2-x}Sr_xCuO_4$ (b) & $Bi_2(Sr,La)_2CuO_y$ (a)&$Y_{0.9}Ca_{0.1} Ba_2 Cu_3O_y$ (a)\\
				\hline
				$T_c~(K)$ & 38 & 28.3 &80.6 \\
				$\xi~(a_B)$ & 65 & 76 &64 \\
				$\Delta~(E_H)$ & 0.000212 & 0.000158 &0.00045 \\
				$E_F~(E_H)$ & 0.000935 & 0.000701 &0.00411 \\
				${\delta}E~(E_H)$ & 0.000356& 0.000267 &0.00156 \\
				$L_y~(a_B)$ & 120 & 139 &57 \\
				$T_0~(K)$ & 8.13 & 6.1 &35.8 \\
				\hline
				\hline
			\end{tabular}
		}
		\caption{Cuprate materials with coherence length of order of the width of the nanowire. The temperature $T_0$ allows the Fermi distribution to spread over the entire energy ${\delta}E$. (Based on data for $T_c$ and $\xi$ from (a) \cite{wesche2015physical},  Table 9.1 and (b) \cite{Cyrot}, Table 7.4. }
		\label{tab:3bump}
	\end{center}
\end{table}
\section{Current and Shot Noise in the Left Lead}
\label{sec:shot noise}

The average current is obtained by taking the thermal average of the current operator so that
\begin{eqnarray}
{\langle} J_L(z,t) {\rangle}=\frac{{\hbar}{\rm e}}{2im_L}{\bigg{\langle}}{\psi}_L^{\dagger}(z,t)\frac{d{\psi}_L(z,t)}{dz}-\frac{d{\psi}_L^{\dagger}(z,t)}{dz}{\psi}_L(z,t) {\bigg {\rangle}},
\end{eqnarray}
where ${\psi}_L(z,t)$ is given by Eq. (\ref{field1}). 
The average current,  in terms of scattering amplitudes, is then given by
\begin{equation}
\begin{aligned}
{\langle}J_L{\rangle} &=\frac{{\rm e}}{h}{\int}de~{\bigg(}F^L_h~( |r^{LL}_{hh}|^2-1) ~ -N^L_p( |r^{LL}_{pp}|^2-1)~\\
& +F^R_h~|t^{LR}_{hh}|^2 +F^L_p~|r^{LL}_{hp}|^2 +F^R_p~|t^{LR}_{hp}|^2 ~ \\
& -N^L_h~|r^{LL}_{ph}|^2 -N^R_h ~|t^{LR}_{ph}|^2 -N^R_p~|t^{LR}_{pp}|^2 {\bigg)},
 \end{aligned}
\end{equation}
where   $N^L_p={\langle}{\hat a}^{\dagger}_{pL}{\hat a}_{pL} {\rangle}$,  $F^L_p={\langle}{\hat a}_{pL}{\hat a}^{\dagger}_{pL} {\rangle}$, $N^L_h={\langle}{\hat a}^{\dagger}_{hL}{\hat a}_{hL} {\rangle}$,  $F^L_h={\langle}{\hat a}_{hL}{\hat a}^{\dagger}_{hL} {\rangle}$.  
The average current is zero if the Fermi energies in the left and right reservoirs are equal. However, if the chemical potentials are different in the two reservoirs, there will be a net current flow.  

%

The shot noise in the left lead is given by  correlations of fluctuations about the average current in the left lead at different times,
\begin{eqnarray}
S^{LL}(z,y;t,s)={\langle} (J_L(z,t)-{\langle}J_L(z,t){\rangle})(J_L(y,s)-{\langle}J_L(y,s){\rangle}) {\rangle}
\end{eqnarray}
We call this the ``local" shot noise because it only involves current fluctuations in the left lead. 

If only the zero frequency limit of this correlation function is kept, we obtain the zero frequency component of the local shot noise
\begin{equation}
	\begin{aligned}
	S^{LL}({\omega}=0) ={\rm S}^{LL}_{pp}+{\rm S}^{LL}_{hh}+{\rm S}^{LL}_{ph}.	\end{aligned}
\label{localshot1}	
\end{equation}
In Eq. (\ref{localshot1}), we have separated the shot noise into components  ${\rm S}^{LL}_{pp}$ and ${\rm S}^{LL}_{hh}$, which give the contribution to the shot noise  due only to particles or to holes, respectively,  in the left lead, and a contribution ${\rm S}^{LL}_{ph}$ which gives the contribution to the noise due to correlations between particle and hole fluctuations in the left lead. 
These three types of contributions to the local shot noise  can be computed explicitly and are given by
\begin{equation}
	\begin{aligned}
	{\rm S}^{LL}_{pp}& =\frac{m_L^2}{{\pi}^2{\hbar}^4}{\int}de ~{\bigg[} -F^L_p~N^L_p ~(1- |r^{LL}_{hp}|^2 - |r^{LL}_{pp}|^2 )^2 \\
	   & -F^R_p ~N^R_p ~( |t^{LR}_{hp}|^2+ |t^{LR}_{pp}|^2  )^2\\
	   & -( F^R_p~N^L_p+F^L_p N^R_p) ~|( r^{LL *}_{pp} t^{LR}_{pp}  -r^{LL}_{hp} t^{LR *}_{hp} )|^2 
	~ {\bigg]}
	\end{aligned}
\end{equation}
\begin{equation}
	\begin{aligned}
	{\rm S}^{LL}_{hh}& =\frac{m_L^2}{{\pi}^2{\hbar}^4}{\int}de ~{\bigg[} -F^L_h~ N^L_h~ (1- |r^{LL}_{hh}|^2 - |r^{LL}_{ph}|^2)^2  \\
	   & - F^R_h~N^R_h ~( |t^{LR}_{hh}|^2 + |t^{LR}_{ph}|^2)^2 \\
	   &  -( F^R_h~N^L_h+F^L_h N^R_h)~| ( r^{LL *}_{ph} t^{LR}_{ph} - r^{LL}_{hh}  t^{LR *}_{hh})|^2 
	~ {\bigg]}
	\end{aligned}
\end{equation}
\begin{equation}
	\begin{aligned}
	{\rm S}^{LL}_{ph}& =\frac{m_L^2}{{\pi}^2{\hbar}^4}{\int}de ~{\bigg[} 
	- (F^L_p~N^L_h +F^L_h~N^L_p)~| ( r^{LL}_{hh} r^{LL *}_{hp}- r^{LL}_{pp} r^{LL *}_{ph}) |^2~\\
	&  -(F^R_p~ N^R_h+F^R_h~ N^R_p) ~|( t^{LR}_{hh} t^{LR *}_{hp}  - t^{LR}_{pp} t^{LR *}_{ph}  )|^2 \\
	   & -(F^R_p~ N^L_h +F^L_hN^R_p)~|( r^{LL *}_{ph} t^{LR}_{pp} - r^{LL}_{hh} t^{LR *}_{hp} )|^2\\
	   & -(F^L_p~N^R_h +F^R_h N^L_p)~|( r^{LL *}_{pp} t^{LR}_{ph}-r^{LL}_{hp} t^{LR *}_{hh}  )|^2
	~ {\bigg]}
	\end{aligned}
\end{equation}

We now consider the local shot noise for a LSCO nanowire ($La_{2-x}Sr_xCuO_4$ in Table I), for the case when the chemical potentials in the left and right reservoirs differ by $V=0.00002$. We consider two lengths  of the superconducting segment  $L_S=5000$ and $L_S=500$, and we assume that the temperature of the nanowire is $T=8.13~K$. 
The  Fermi distribution, $N^L_p$ for particles at temperature $T=8.13~K$ is plotted in Fig. 3.  The Fermi distribution for holes $N^L_h$ is  identical. However, $N^L_p$  and $N^L_h$ measure the distributions of particles and holes, on their respective sides of the Fermi surface, as a function of distance in energy from the Fermi surface.

In  Fig. \ref{fig:total_diff_shot_noise_local_angle_resolved}, we show  contour plots of the differential shot noise, $dS{\equiv}dS^{LL}/de$ in the left lead, as a function of gap angle $\theta$ and energy.  
Fig. \ref{fig:total_diff_shot_noise_local_angle_resolved}.a (Fig. \ref{fig:total_diff_shot_noise_local_angle_resolved}.b) is for a superconducting segment of length $L_S=5000$ ($L_S=500$).  Figs. \ref{fig:total_diff_shot_noise_slices}.a and \ref{fig:total_diff_shot_noise_slices}.b show cuts through the Fig. \ref{fig:total_diff_shot_noise_local_angle_resolved} contour plots at angles $\theta=0$ and $\theta=\pi/5$. In all cases, the shot noise is negative which is a signature of electron anti-bunching due to their Fermion statistics \cite{Henny,Oliver}.

  For energies below the gap energy ${\Delta}=0.000212$, the shot noise is non-zero and decreases in  magnitude with increasing energy.  This contribution to the shot noise is due to particles and holes in the left lead that are involved with Andreev reflection from the superconductor.  The decreasing magnitude of the noise is due to the decrease in the particle and hole Fermi distributions with increasing energy.  With increasing energy, there are fewer particles and holes available to contribute to the shot noise. For energies above the gap, and for  $L_S=5000$ , there are numerous peaks in the magnitude of the noise.  These are due to scattering resonances inside the superconductor for energies above the gap, which allows selective transmission of particles and holes into and out of the superconductor. The magnitude of the shot noise drops to zero as energies approach $e=0.00035$ because the Fermi distributions tends to zero at those energies.

\section{Cross Correlated Shot Noise}
\label{sec:nonlocal shot noise}

Cooper pair splitting (CPS) requires spatial separation of entangled electron pairs formed by the cooper pairing mechanism in superconductors \cite{lesovik}. Therefore, the efficiency of Cooper pair splitting can be related to the cross correlated ({\it nonlocal}) shot noise between different leads in the superconducting device \cite{Celis,buttikersign}.
A number of geometries have been considered for these devices, including the linear geometry used in this paper. Cooper pair splitting dynamics are typically governed by the relationship between the nonlocal scattering processes of the device, such as crossed Andreev reflection (CAR) which correspond to $p\rightarrow h$ and $h\rightarrow p$ transmission and elastic cotunneling (EC), which corresponds to  $p\rightarrow p$ and $h\rightarrow h$ transmission. In general, the CAR process is known to be key process for the functioning of CPS (though not necessarily the main source of positive noise cross-correlations in the high transparency regime \cite{Freyn,buttikersign}) and it's maximization is a key goal of CPS design. 

We can compute the correlations between current fluctuations in the left and right leads.   This ``cross-correlated" (nonlocal) shot noise is defined as
\begin{eqnarray}
S^{LR}(z,y;t,s)=\frac{1}{2}{\big \langle} (J_L(z,t)-{\langle}J_L{\rangle})(J_R(y,s)-{\langle}J_R{\rangle}) ~~~~~ \nonumber\\
+(J_R(y,s)-{\langle}J_R{\rangle})(J_L(z,t)-{\langle}J_L{\rangle}) {\big \rangle}.
\end{eqnarray}
where the current $J_R$ is constructed from right-lead field operators ${\hat \psi}_R(z,t)$ which, in turn, are defined in a manner analogous to that given in Eq. (\ref{field1}) for the left lead. 
In the zero frequency limit we obtain the cross-correlated shot noise
\begin{equation}
	S^{LR}({\omega}=0) ~{\equiv}~{\rm S}^{LR}_{pp}+{\rm S}^{LR}_{hh}+{\rm S}^{LR}_{ph},	
	\label{eqn:nonlocal_noise_decomp}
\end{equation}
where, again,  we have separated the shot noise into components  ${\rm S}^{LR}_{pp}$,  ${\rm S}^{LR}_{hh}$ and ${\rm S}^{LR}_{ph}$. The first two terms of Eq. (\ref{eqn:nonlocal_noise_decomp}), ${\rm S}^{LR}_{pp}$ and ${\rm S}^{LR}_{hh}$ give the contributions  due to correlations between particles in the left and right leads and between holes in the left and right leads, respectively. The final term gives the the contribution due to correlations between particles and holes in the left and right leads. These correlation functions can be calculated explicitly and the expressions for the three components of Eq.  (\ref{eqn:nonlocal_noise_decomp}) are given by
\begin{equation}
\begin{aligned}
	{\rm S}^{LR}_{hh} =&\frac{ m_L m_R}{\pi ^2{\hbar}^4}~{\int}de ~{\bigg[} F^L_hN^L_h {\big(}  (|r^{LL}_{hh}|^2+|r^{LL}_{ph}|^2-1)(|t^{RL}_{hh}|^2  +|t^{RL}_{ph}|^2) {\big)}\\
   &+F^R_h N^R_h{\big(} (|r^{RR}_{hh}|^2+|r^{RR}_{ph}|^2-1)(|t^{LR}_{hh}|^2+|t^{LR}_{ph}|^2)  {\big)}\\
   &+(F^L_h N^R_h+F^R_h N^L_h)~{\big(} {\rm Re}\left[r^{LL}_{hh}r^{RR}_{hh} t^{LR*}_{hh} t^{RL*}_{hh}\right] -{\rm Re}\left[r^{LL}_{hh}r^{RR*}_{ph}t^{RL}_{ph}t^{LR*}_{hh}\right]\\
   &+{\rm Re} \left[r^{LL}_{ph} r^{RR}_{ph}t^{LR*}_{ph} t^{RL*}_{ph} \right]  -{\rm Re}\left[r^{LL}_{ph}r^{RR*}_{hh}t^{RL}_{hh}t^{LR*}_{ph}\right]  {\big)}	~ {\bigg]}
\end{aligned}
\end{equation}
\begin{equation}
\begin{aligned}
	{\rm S}^{LR}_{pp} &= \frac{ m_L m_R}{\pi ^2{\hbar}^4}~{\int}de {\bigg[}F^L_p N^L_p {\big(} (|r^{LL}_{hp}|^2+|r^{LL}_{pp}|^2 -1)~ (|t^{RL}_{hp}|^2  +|t^{RL}_{pp}|^2)  {\big)}\\
   & + F^R_p N^R_p {\big(}(|r^{RR}_{hp}|^2+|r^{RR}_{pp}|^2 -1)  (|t^{LR}_{hp}|^2 +|t^{LR}_{pp}|^2)  {\big)}\\
   &+ ( F^L_p N^R_p+F^R_p N^L_p) {\big(} {\rm Re}\left[ r^{LL}_{hp} r^{RR}_{hp} t^{LR*}_{hp} t^{RL*}_{hp} \right] - {\rm Re}\left[r^{LL}_{hp} r^{RR*}_{pp} t^{RL}_{pp}t^{LR*}_{hp} \right]\\
   & + {\rm Re} \left[r^{LL}_{pp} r^{RR}_{pp} t^{LR*}_{pp}  t^{RL*}_{pp}\right] - {\rm Re} \left[ r^{LL}_{pp} r^{RR*}_{hp} t^{RL}_{hp}t^{LR*}_{pp} \right]  {\big)}{\bigg]}
   \end{aligned}
   \end{equation}
\begin{equation}
\begin{aligned}
	{\rm S}^{LR}_{ph} = & \frac{ m_L m_R}{\pi^2{\hbar}^4}~{\int}de {\bigg[} (~F^L_p~N^L_h+F^L_h~N^L_p) {\big(} {\rm Re} \left[ r^{LL}_{hh}  r^{LL*}_{hp} t^{RL}_{hp}t^{RL*}_{hh} \right]
    - {\rm Re} \left[ r^{LL}_{hh} r^{LL*}_{hp}  t^{RL}_{ph} t^{RL*}_{pp}\right]\\
   &-{ \rm Re} \left[r^{LL}_{ph} r^{LL*}_{pp}t^{RL}_{hh} t^{RL*}_{hp}\right] + {\rm Re} \left[ r^{LL}_{ph} r^{LL*}_{pp} t^{RL}_{pp}  t^{RL*}_{ph}~\right]{\big)}\\
   &+(F^R_p~ N^R_h+F^R_h~N^R_p)~~{\big(} {\rm Re} \left[ r^{RR}_{hh} r^{RR*}_{hp}t^{LR}_{hp} t^{LR*}_{hh}\right]- {\rm Re} \left[ r^{RR}_{hh} r^{RR*}_{hp}  t^{LR}_{ph}t^{LR*}_{pp}\right]\\
   &- {\rm Re} \left[ r^{RR}_{ph} r^{RR*}_{pp} t^{LR}_{hh} t^{LR*}_{hp}\right] + {\rm Re} \left[ r^{RR}_{ph} r^{RR*}_{pp}t^{LR}_{pp} t^{LR*}_{ph}~\right]~   {\big)}\\
   &+(F^L_h~N^R_p+F^R_p~ N^L_h)~{\big(} {\rm Re} \left[ r^{LL}_{hh} r^{RR}_{hp} t^{LR*}_{hp} t^{RL*}_{hh} \right]- {\rm Re} \left[ r^{LL}_{hh} r^{RR*}_{pp}  t^{RL}_{ph}t^{LR*}_{hp}\right]\\
   &+{\rm Re} \left[ r^{LL}_{ph}r^{RR}_{pp} t^{LR*}_{pp} t^{RL*}_{ph} \right] - {\rm Re} \left[ r^{LL}_{ph} r^{RR*}_{hp} t^{RL}_{hh}t^{LR*}_{pp}\right] {\big)}\\
   &+(F^L_p~N^R_h  + F^R_h~ N^L_p )~{\big(} {\rm Re} \left[ r^{LL}_{hp}r^{RR}_{hh}t^{LR*}_{hh}t^{RL*}_{hp} \right] - {\rm Re} \left[r^{LL}_{hp} r^{RR*}_{ph}  t^{RL}_{pp}t^{LR*}_{hh}~\right]\\
   &+{\rm Re} \left[r^{LL}_{pp}r^{RR}_{ph} t^{LR*}_{ph}  t^{RL*}_{pp} \right] - {\rm Re} \left[ r^{LL}_{pp} r^{RR*}_{hh} t^{RL}_{hp} t^{LR*}_{ph}\right]{\big)}~{\bigg]} {\bigg]}
 \end{aligned}
\end{equation}

We start our investigation of the behavior of these expressions in the HTS parameter regime by evaluating $dS^{LR}{\equiv}dS^{LR}/de$  as a function of the orientation of the order parameter $\theta$ and the energy. In Fig. \ref{fig:nonlocal_noise_angle_resolved_plots} we show contour plots for $dS^{LR}$ over the range of angles $\theta=[-\pi/4,\pi/4]$ and for energies up to the edge of the first transverse mode. This is done for two different lengths for the superconducting region, $L_S$, with Fig. \ref{fig:nonlocal_noise_angle_resolved_plots}.a corresponding to $L_S=5000$ and \ref{fig:nonlocal_noise_angle_resolved_plots}.b to $L_S=500$. In addition, we bias the junction asymmetrically by setting the potential in the left lead $V_L=.00002$ and by grounding the superconductor and right lead ($V_S=V_R=0$) , and we biased it symmetrically by setting $V_L=V_R=.00002$ and grounding the superconductor. In Fig. \ref{fig:nonlocal_shot_noise_slices}, we give the plots of $dS^{LR}$ corresponding to cross-sectional slices with $\theta= 0,\pi/5$, as labeled in Fig. \ref{fig:nonlocal_noise_angle_resolved_plots}. Here Fig.  \ref{fig:nonlocal_shot_noise_slices}.a contains the $\theta=0$ slice for both values of $L_S$ and  Fig. \ref{fig:nonlocal_shot_noise_slices}.b contains the $\theta=\pi/5$ slices for both lengths.

The nonlocal shot noise can be partitioned into components for $S_{pp}$, $S_{hh}$ and $S_{ph}$. In Figs. \ref{fig:nonlocal_shot_noise_slices_comp} and \ref{fig:nonlocal_shot_noise_slices_comp_pi5} plots of $dS_{hh}$ and $dS_{ph}$ are shown with Fig. \ref{fig:nonlocal_shot_noise_slices_comp} corresponding to the $\theta=0$ slices and Fig. \ref{fig:nonlocal_shot_noise_slices_comp_pi5} to the $\theta=\pi/5$ slices. One key feature we immediately notice is that while the $S^{LR}_{hh}$ and $S^{LR}_{pp}$ terms are always negative for all energies, the $S^{LR}_{ph}$ term is positive for some regions of the energy range. Positive values of the nonlocal noise indicate electron bunching and are often associated with the generation of entanglement in CPS devices. The plots in Figs. \ref{fig:nonlocal_shot_noise_slices_comp} and \ref{fig:nonlocal_shot_noise_slices_comp_pi5} are given for both asymmetric ($V_L=V,V_R=V_S=0$) and symmetric biasing ($V_L=V_R=V,V_S=0$) of the junction. While the two different biasing schemes differ somewhat quantitatively, share the same qualitative behavior.

Also of interest is how the nonlocal noise changes as a function of the length of the superconducting region. In Fig. \ref{fig:nonlocal_shot_noise_length_behavior} we plot the total nonlocal noise as well as the $h\rightarrow h$ and $p\rightarrow h$ contributions versus the length of the superconducting region with respect to a fixed value of $V_L=.00002, V_R=V_S=0$. What we find for the total nonlocal short noise is that it monotonically increases, approaching zero from below as the system size increases; likewise for the $h \rightarrow h$ and $p \rightarrow p$ contributions. The $p \rightarrow h$ contribution, however, displays a non-monotonic behavior, with a minimum at $L_S\approx 250$ before approaching zero again.

\section{Conclusions}
\label{sec:conclude}

Exact expressions for the scattering matrix of a nanowire NSN junction have been found using scattering theory. These expressions correspond to the high transparency limit and are done without the Andreev approximation. Likewise, exact expressions for the local and nonlocal shot noise associated with current propagation in the nanowire were found. These exact expressions for the current and the shot noise are functions of the length of the superconducting segment of the wire, the temperatures and Fermi energies of the electron reservoirs, the effective mass of the particles in the three segments of the nanowire, and the superconducting gap.  The expressions are general enough that they can be applied to current and shot noise calculations performed using the fairly standard quasi-1d approximation for any NSN wire.

We have  determined the smallest size nanowire that allows the transmission of a superconducting current through a nanowire, based on the condition that the width of the wire be of order of the coherence length of the superconducting material using values of the critical temperature $T_c$ and the coherence length ${\xi}$ for common high temperature superconductors listed in Table 1.  We found that the most favorable materials are $La_{1.45}Nd_{0.4}Sr_{0.15}CuO_{4+{\delta}}$, $Nd_{1.85}Ce_{0.15}CuO_{4-{\delta}}$, $Bi_2(Sr,La)_2CuO_y$, and $Y_{0.9}Ca_{0.1} Ba_2 Cu_3O_y$. In all four cases, the coherence length of the superconductor appears to be shorter than the width of the nanowire that allows current propagation in the first propagating channel of the nanowire. Recent  developments in materials science have shown that fabrication of structures with dimensions on the order of those found from our model is now on the cusp of what is possible experimentally. 

Using realistic parameters for LSCO,  we have investigated the local and nonlocal shot noise properties of the NSN structure.  We have found that the local shot noise is always negative, as is expected because of anti-bunching due to Fermi statistics of the electrons.  However, the nonlocal shot noise shows electron bunching (regions of positive shot noise) due to the correlations (entanglement) between electrons and holes that emerge from opposite ends of the superconducting segment due to break-up of Cooper pairs inside the superconductor. Nonlocal shot noise in the NSN structure can, therefore, provide a measure of the effectiveness of entangled electron pair production in the nanowire.

\vspace{0.3cm}

\section*{Acknowledgments} The authors thank the Robert A. Welch Foundation (Grant No. F-1051) for support of this work.  

\section{Appendix A: Thermal Distribution of Type A and B Electrons}

Consider the electrons that enter the waveguide from the left thermal reservoir, which is assumed  to be a low temperature Fermi gas.  The density operator for electrons in the left thermal reservoir is given by 
${\hat \rho}_{L}=\frac{{\rm e}^{-{\beta}{\hat  K}_{L}}}{{\rm Tr}[{\rm e}^{-{\beta}{\hat  K}_{L}}]}$
where ${\hat  K}_{L}={\int}dE(E-E^{L}_F){\hat a}(E)^{\dagger}_{L}{\hat a}(E)_{{L}}$ and $E_F^{L}$ is the Fermi energy of the left  reservoir.  Using Wick's theorem we can write
\begin{eqnarray}
{\langle} {\hat a}(E)_L^{\dagger}{\hat a}(E')_L {\rangle}=   {\rm Tr}[{\hat \rho}_{L} {\hat a}(E)^{\dagger}_{L}{\hat a}(E')_{{L}} ]={\delta}(E-E') f^{L}(E)
\end{eqnarray}
where 
$ f^{L}(E)=\left(1+{\rm e}^{{\beta}(E-E_F^{L})} \right)^{-1}$ 
is the Fermi distribution of electrons in the left thermal reservoir.  
Now consider some function of energy, $G(E)$. We assume that the temperature is low enough that the distribution $f^L(E)=1$  for $E<E_{1,1}^{tr}$ and $f^L(E)=0$ for $E>E_{2,1}$. Thus
\begin{eqnarray}
{\langle}G{\rangle}_L=C_1+{\int_{E_{1,1}^{tr}}^{E_{2,1}^{tr}}}dE~\frac{G(E)}{1+{\rm e}^{{\beta}(E-E_F^{L})} }.
\end{eqnarray}
where $C_1={\int_{0}^{E_{1,1}^{tr}}}dE~G(E)$ is a constant independent of temperature.  We choose $E^L_F=(E_{1,1}^{tr}+E_{2,1}^{tr})/2$. Then let $E=e+E_{1,1}^{tr}$ so $E-E^L_F=e+E_{1,1}^{tr}-E^L_F=e-{\epsilon}^L_F$, where ${\epsilon}^L_F=(E_{2,1}^{tr}-E_{1,1}^{tr})/2$. The average ${\langle}G{\rangle}_{L}$,  can now be written
\begin{eqnarray}
{\langle}G{\rangle}_{L}=C_1+{\int_{0}^{2{\epsilon}^{L}_F}} de~\frac{G'(e)}{1+{\rm e}^{{\beta}(e-{\epsilon}^{L}_F)} }.
\end{eqnarray}
where $G'(e)=G(e+E_{1,1}^{tr})$. 

Next write the thermal average in terms of A-particle and  B-particle energies \cite{Reichl}. We can change variables so $e={\epsilon}^{L}_F-e^{L}_{B}$ with $0{\leq}e^{L}_{B}{\leq}{\epsilon}^L_F$, and  $e=e^{L}_{A}-{\epsilon}^{L}_F$ with ${\epsilon}^{L}_F{\leq}e^{L}_{A}{\leq}2{\epsilon}^{L}_F$.   Then the thermal average can be written
\begin{equation}
\begin{aligned}
{\langle}G{\rangle} &=C_1+ {\int_{0}^{{\epsilon}_F}} de^{L}_B~~\frac{G'(e^{L}_B)}{1+{\rm e}^{{\beta}(e^{L}_B-{\epsilon}_F)} }.~+{\int_{{\epsilon}_F}^{2{\epsilon}_F}} de^{L}_{A}~~\frac{G'(e^{L}_{A})}{1+{\rm e}^{{\beta}(e^{L}_{A}-{\epsilon}_F)} }.\\
&=C_1+{\int_{0}^{{\epsilon}_F}} d{e}~\frac{{\rm G}_B(e)}{(1+{\rm e}^{-{\beta}{ e}})}~+{\int_{0}^{{\epsilon}_F}} d{e}~\frac{{\rm G}_A({ e})}{(1+{\rm e}^{{\beta}{e}})}~~
\end{aligned}
\end{equation}
where the thermal distributions are 
$f_A({e})=\frac{1}{(1+{\rm e}^{{\beta}{ e}})}$  and $ f_B({ e})=\frac{1}{(1+{\rm e}^{-{\beta}{ e}})}$, respectively, and describe the distribution of particles for energy moving away from the Fermi energy.

\section{Appendix B: Transmission and Reflection Amplitudes}
\begin{equation}
\begin{aligned}
r^{LL}_{pp} &=-\frac{1}{\text{DEN}} {\bigg[} 8 k_h k_p u_o^2 v_o^2 (q^L_h-q^L_p) (q^R_h+q^R_p) ~\\
&+e^{i L_S k_h-i L_Sk_p} \left(u_o^2 P^L_{hh} M^L_{pp}-v_o^2 P^L_{hp}  M^L_{ph}\right) (u_o^2 P^R_{hh}P^R_{pp}-v_o^2   M^R_{hp} M^R_{ph})\\
& +   e^{i L_S k_h+i L_S k_p} \left(\text{vo}^2 P^L_{hp} P^L_{ph}-u_o^2 P^L_{hh} P^L_{pp}\right)  \left(u_o^2 P^R_{hh} M^R_{pp}-v_o^2 M^R_{hp} P^R_{ph}\right) \\
& +  e^{-i L_S k_h-iL_S  k_p} \left(v_o^2  M^L_{hp}  M^L_{ph}-u_o^2 M^L_{hh}  M^L_{pp}\right)  \left(u_o^2 M^R_{hh} P^R_{pp}-v_o^2 P^R_{hp} M^R_{ph}\right) \\
&  +e^{iL_S k_p-i L_S k_h} \left(u_o^2 M^L_{hh} P^L_{pp}-v_o^2 M^L_{hp} P^L_{ph}\right)   \left(u_o^2 M^R_{hh}
  M^R_{pp}-v_o^2  P^R_{hp} P^R_{ph}\right) {\bigg]}
\end{aligned}
\end{equation}
 where $P^X_{y,z}=k_y+q^X_z$ and $M^X_{y,z}=k_y-q^X_z$, where $X={L,R}$, $y={p,h}$ and $z={p,h}$.  For example, $P^L_{p,h}=k_p+q^L_h$ and $M^R_{p,p}=k_p-q^R_p$. 

\begin{equation}
\begin{aligned}
r^{LL}_{ph} &= \frac{2u_ov_o}{\text{DEN}} \sqrt{q^L_h} \sqrt{q^L_p}{\bigg[}~4k_hk_p (q^R_h+q^R_p)
   \left(u_o^2+v_o^2\right) \\
  & +(k_h +k_p) e^{i L_Sk_p-i L_Sk_h} \left( u_o^2 M^R_{hh} M^R_{pp} -v_o^2 P^R_{hp} P^R_{ph}\right) \\
  & +  (k_h-k_p) e^{-i  L_S k_h-i L_S k_p} \left(u_o^2 M^R_{hh} P^R_{pp}-v_o^2 P^R_{hp} M^R_{ph}\right) \\
  &  - (k_h+k_p)   e^{i L_Sk_h-i L_S k_p} \left(u_o^2 P^R_{hh} P^R_{pp}-v_o^2 M^R_{hp}M^R_{ph}\right) \\
  &  -  (k_h-k_p)  e^{iL_S k_h+i L_Sk_p} \left(u_o^2 P^R_{hh}M^R_{pp} -v_o^2  M^R_{hp} P^R_{ph}\right)  {\bigg]}
   \end{aligned}
\end{equation}
\begin{equation}
\begin{aligned}
r^{LL}_{hp} & = \frac{u_o v_o}{\text{DEN}} \sqrt{q^L_h} \sqrt{q^L_p}  {\bigg[}8 k_hk_p (q^R_h+q^R_p)
   \left(u_o^2+v_o^2\right) \\
 & -2   (k_h-k_p) e^{iL_Sk_h+i L_Sk_p} \left(u_o^2 P^R_{hh}M^R_{pp}-v_o^2 M^R_{hp} P^R_{ph}\right)\\
 &  +2  (k_h+k_p) e^{i L_Sk_p-i L_Sk_h} \left(u_o^2 M^R_{hh}M^R_{pp}-v_o^2  P^R_{hp} P^R_{ph}\right) \\
 &  -2  (k_h+k_p) e^{iL_S k_h-i L_Sk_p} \left(u_o^2 P^R_{hh} P^R_{pp}-v_o^2 M^R_{hp} M^R_{ph}\right) \\
 &  + 2 (k_h-k_p) e^{-iL_Sk_h-i L_S k_p} \left(u_o^2 M^R_{hh} P^R_{pp}-v_o^2  P^R_{hp}  M^R_{ph} \right){\bigg]}
 \end{aligned}
\end{equation}
\begin{equation}
\begin{aligned}
r^{LL}_{hh} &=\frac{1}{\text{DEN}} {\bigg[}~8 k_h k_p u_o^2 v_o^2 (q^L_h-q^L_p)(q^R_h+q^R_p)\\
& +e^{i L_Sk_h+i L_S k_p} \left(v_o^2 M^L_{hp}  M^L_{ph}-u_o^2  M^L_{hh}  M^L_{pp} \right)  \left(u_o^2 P^R_{hh}
   M^R_{pp}-v_o^2  M^R_{hp}   P^R_{ph}\right)~ \\
& +e^{iL_S k_p-i L_S  k_h} \left(u_o^2P^L_{hh}M^L_{pp}   -v_o^2 P^L_{hp}  M^L_{ph}\right)   \left(u_o^2 M^R_{hh} M^R_{pp}-v_o^2 P^R_{hp} P^R_{ph}\right)~\\
&+e^{iL_S k_h-i L_S  k_p} \left(u_o^2M^L_{hh}P^L_{pp}  -v_o^2 M^L_{hp} P^L_{ph}\right) 
 \left(u_o^2 P^R_{hh} P^R_{pp} -v_o^2 M^R_{hp} M^R_{ph}\right)~\\
&+e^{-iL_S k_h-i L_S  k_p} \left(v_o^2 P^L_{hp} P^L_{ph}-u_o^2 P^L_{hh}P^L_{pp}\right) \left(u_o^2 M^R_{hh} P^R_{pp}-v_o^2 P^R_{hp} M^R_{ph}\right) {\bigg]}~\\
    \end{aligned}
\end{equation}
\begin{equation}
	\begin{aligned}
	t^{RL}_{pp} & =\frac{4}{\text{DEN}}~  \left(u_o^2-v_o^2\right) \sqrt{q^L_p} \sqrt{q^R_p} \\
	& {\times}{\bigg(} e^{i L_Sk_p-i L_Sq^R_p}~ k_h~  v_o^2P^L_{ph} P^R_{ph}~ -e^{i L_S k_h-iL_Sq^R_p}~k_p~
	u_o^2 P^L_{hh} P^R_{hh}~ \\
	& + e^{-iL_S k_h-i L_Sq^R_p}~ k_p~u_o^2 M^L_{hh} M^R_{hh}-e^{-i L_S k_p-i L_S q^R_p}~ k_h~ v_o^2 M^L_{ph}  M^R_{ph}{\bigg)}\\
	\end{aligned}
\end{equation}
\begin{equation}
	\begin{aligned}
	t^{RL}_{ph} & =\frac{4}{\text{DEN}} ~ \sqrt{q^L_h} \sqrt{q^R_p}~u_o v_o \left(u_o^2-v_o^2\right)~\\
	&{\bigg(}-e^{i L_Sk_h-iL_S q^R_p} ~ k_p ~M^L_{hp} P^R_{hh}~	 + e^{i L_S k_p-i L_S  q^R_p}
	 ~k_h~ M^L_{pp} P^R_{ph}~\\
	& + e^{-i L_S k_h-i L_S q^R_p} ~k_p ~P^L_{hp} M^R_{hh}~ -e^{-iL_S k_p-i L_S  q^R_p}~	k_h~ P^L_{pp}
	  M^R_{ph} {\bigg)} 
	  \end{aligned}
\end{equation}
\begin{equation}
	\begin{aligned}
	t^{RL}_{hp} & =\frac{4}{\text{DEN}} ~ \sqrt{q^L_p}\sqrt{q^R_h} u_o v_o \left(u_o^2-v_o^2\right) \\
	& {\bigg(} -e^{i L_Sk_h+i  L_S q^R_h}~k_p~ P^L_{hh} M^R_{hp} + e^{i L_S q^R_h-i L_S  k_h} ~k_p ~M^L_{hh} P^R_{hp} ~\\
	& +e^{i  L_Sk_p+i L_Sq^R_h}~k_h~ P^L_{ph} M^R_{pp} -e^{iL_S q^R_h-iL_S k_p}~k_h~ M^L_{ph}  P^R_{pp} {\bigg)}
	\end{aligned}
 \end{equation}
\begin{equation}
	\begin{aligned}
	t^{RL}_{hh} & =\frac{4}{\text{DEN}} ~ \sqrt{q^L_h} \sqrt{q^R_h}  \left(u_o^2-v_o^2\right) \\
	 &{\bigg(}-e^{i L_S k_h+i L_S q^R_h}~ k_p~ v_o^2 M^L_{hp} M^R_{hp}~ + e^{i L_S q^R_h-iL_S k_h} ~k_p ~v_o^2
	  P^L_{hp} P^R_{hp}~\\
	  &+e^{i  L_S k_p+i L_S q^R_h}~ k_h~ u_o^2 M^L_{pp}  M^R_{pp} -e^{i L_S q^R_h-iL_S k_p}~ k_h~u_o^2 P^L_{pp}P^R_{pp} {\bigg)} \\
	 \end{aligned}
 \end{equation}

The denominator $DEN$ is given by
\begin{equation}
	\begin{aligned}
	DEN = & 8k_hk_p u_o^2 v_o^2 (q^L_h+q^L_p) (q^R_h+q^R_p)\\
	+ & e^{i L_Sk_p-iL_S k_h} \left(P^L_{hp}P^L_{ph} v_o^2 - M^L_{hh}M^L_{pp} u_o^2 \right) \left( M^R_{hh}
	 M^R_{pp} u_o^2 - P^R_{hp} 
	   P^R_{ph}  v_o^2 \right) \\
	 + & e^{iL_S k_h+i L_S k_p} \left(M^L_{pp}P^L_{hh} u_o^2 - M^L_{hp} P^L_{ph}  v_o^2 \right) \left( M^R_{pp}
	   P^R_{hh} u_o^2 -  M^R_{hp} P^R_{ph}   v_o^2 \right)\\
	   + & e^{iL_Sk_h-i L_S k_p} \left(M^L_{hp} M^L_{ph} v_o^2 -P^L_{hh}P^L_{pp}  u_o^2 \right) \left( P^R_{hh} 
	 P^R_{pp} u_o^2 - M^R_{hp} M^R_{ph} v_o^2 \right)\\
	   + & e^{-i L_S k_h-iL_S k_p} \left(M^L_{hh}P^L_{pp} u_o^2 - M^L_{ph} P^L_{hp} v_o^2 \right) \left(M^R_{hh}
	 P^R_{pp} u_o^2 - M^R_{ph}  P^R_{hp} v_o^2 \right)
	\end{aligned}
\end{equation}

\subsection{List of Figures}

\begin{figure}[!hp]
	\includegraphics[width=\textwidth]{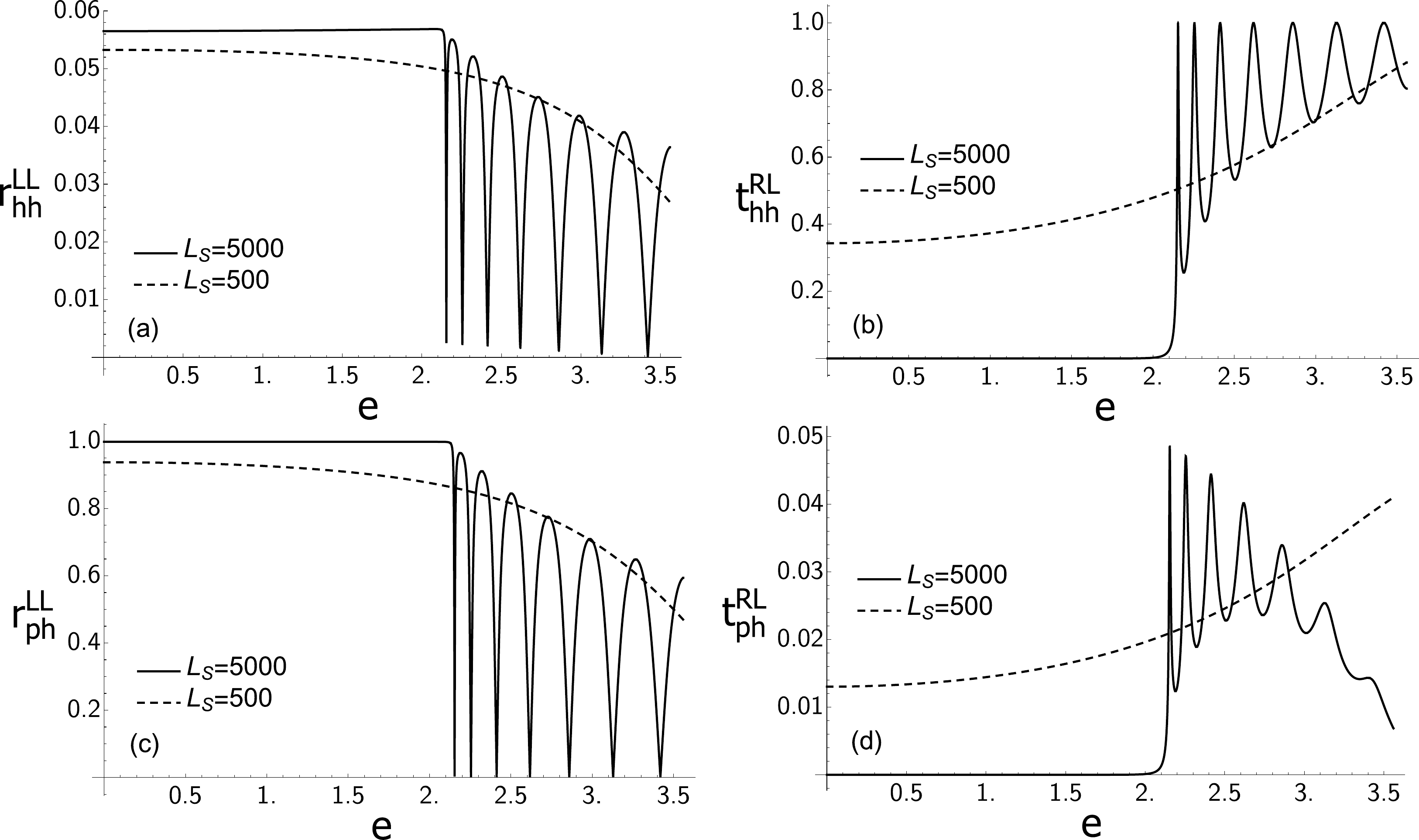}
	\caption{Scattering amplitudes  (a) $r^{LL}_{hh}$ (b) $t^{RL}_{hh}$, (c) $r^{LL}_{ph}$, (d) $t^{RL}_{ph}$,  for a LSCO ($La_{2-x}Sr_xCuO_4$)  nanowire. Because we take the effective mass to be the same in all three parts of the wire,   $r^{LL}_{pp}= r^{LL}_{hh}$, 
		$t^{RL}_{pp}=t^{RL}_{hh}$, $r^{LL}_{hp}=r^{LL}_{ph}$, and $t^{RL}_{hp}=t^{RL}_{ph}$. (In atomic units)}
	\label{fig:1}
\end{figure}
%

\begin{figure}[!hp]
\centering
\includegraphics[width=.5\textwidth]{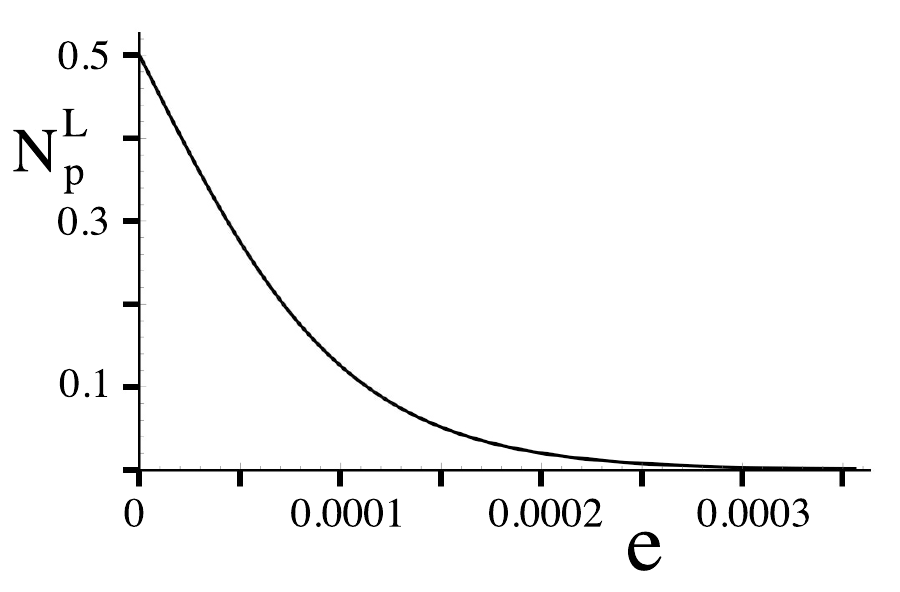}
\caption{ Plot of the Fermi distribution for particles $N^L_p$  for temperature $T=8.13~K$. The Fermi distribution for holes, $N^L_h$, is identical to that of $N^L_p$. Although the distributions $N^L_p$ and $N^L_h$ have the same dependence on energy, they measure particle and hole distributions in terms of  distance in energy from their respective sides of the Fermi surface.  (In atomic units)}
\label{fig:2}
\end{figure}

\begin{figure}[htbp]
	\includegraphics[width=\textwidth]{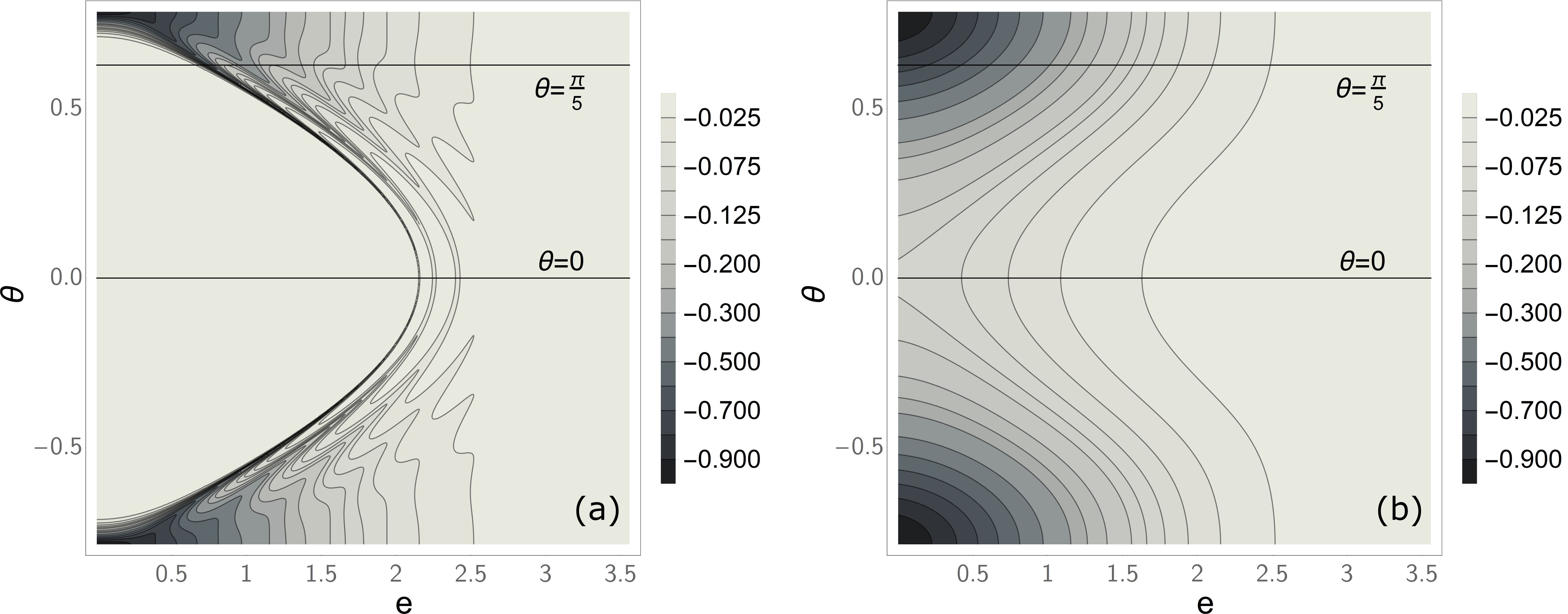}
	\caption{Contour plots of the total local differential shot noise $dS^{LL}$ as a function of energy and the orientation of the order parameter, $\theta$.  (a)  $L_S=5000~au$.  (b)  $L_S=500~au$. (In atomic units)}
	\label{fig:total_diff_shot_noise_local_angle_resolved}
\end{figure}

\begin{figure}[htbp]
	\includegraphics[width=\textwidth]{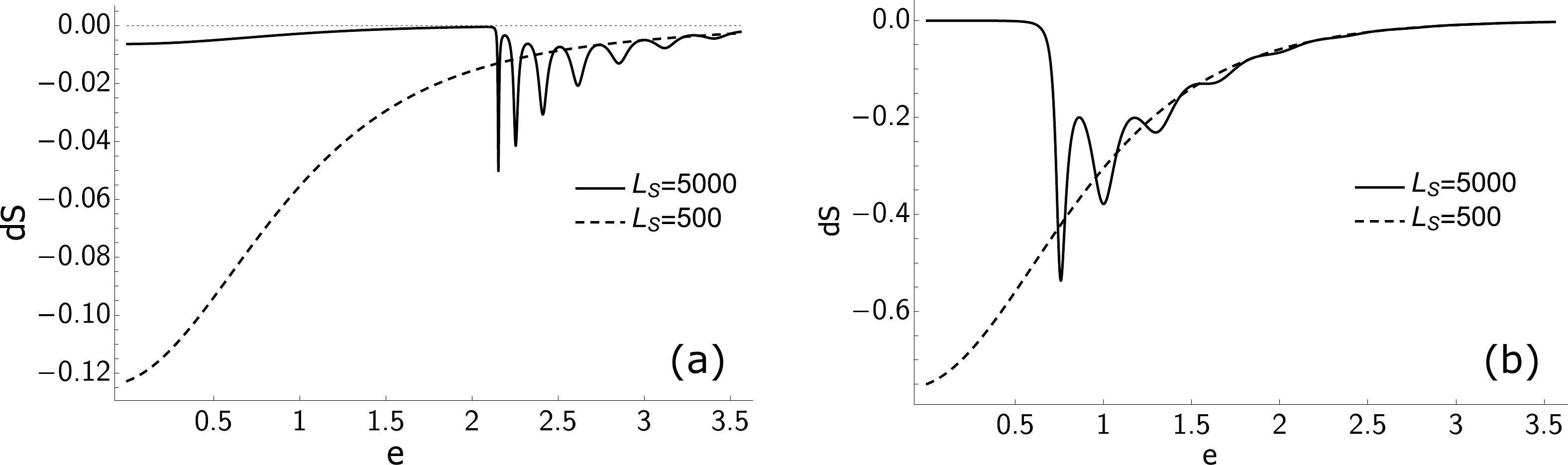}
	\caption{Plots of the total local differential shot noise $dS^{LL}$ as a function of energy. (a) Cross-sectional slice corresponding to $\theta=0$ for $L_S=5000~au$ and in (b)  $L_S=500~au$.  (b) Cross-sectional slice corresponding to $\theta=\pi/5$  for $L_S=5000~au$ and in (b)  $L_S=500~au$. (These slices are marked in Fig. \ref{fig:total_diff_shot_noise_local_angle_resolved}). (In atomic units)}
	\label{fig:total_diff_shot_noise_slices}
\end{figure}

\begin{figure}
	\includegraphics[width=\textwidth]{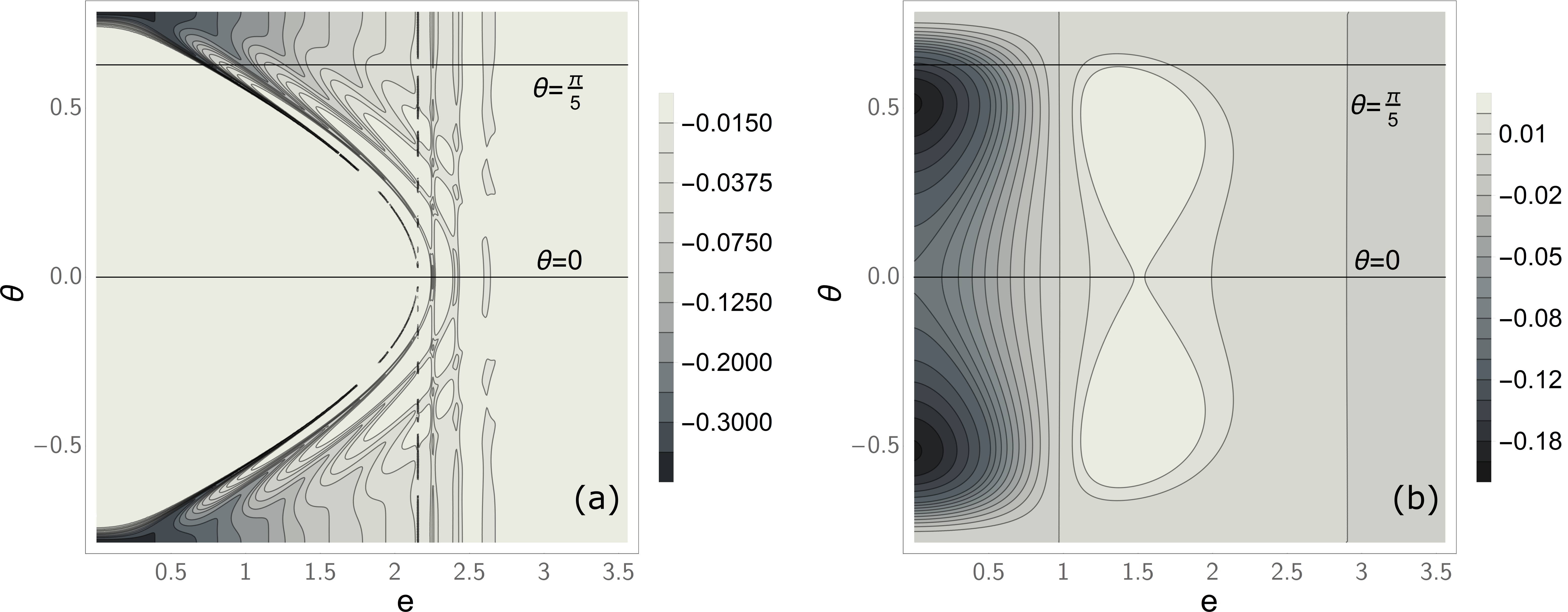}
	\caption{Contour plot of the total differential nonlocal shot noise $dS^{LR}$ as a function of energy and the orientation of the order parameter, $\theta$. (a) $L_S=5000~au$. (b) $L_S=500~au$.  (In atomic units)}
	\label{fig:nonlocal_noise_angle_resolved_plots}
\end{figure}

\begin{figure}[htbp]
	\includegraphics[width=\textwidth]{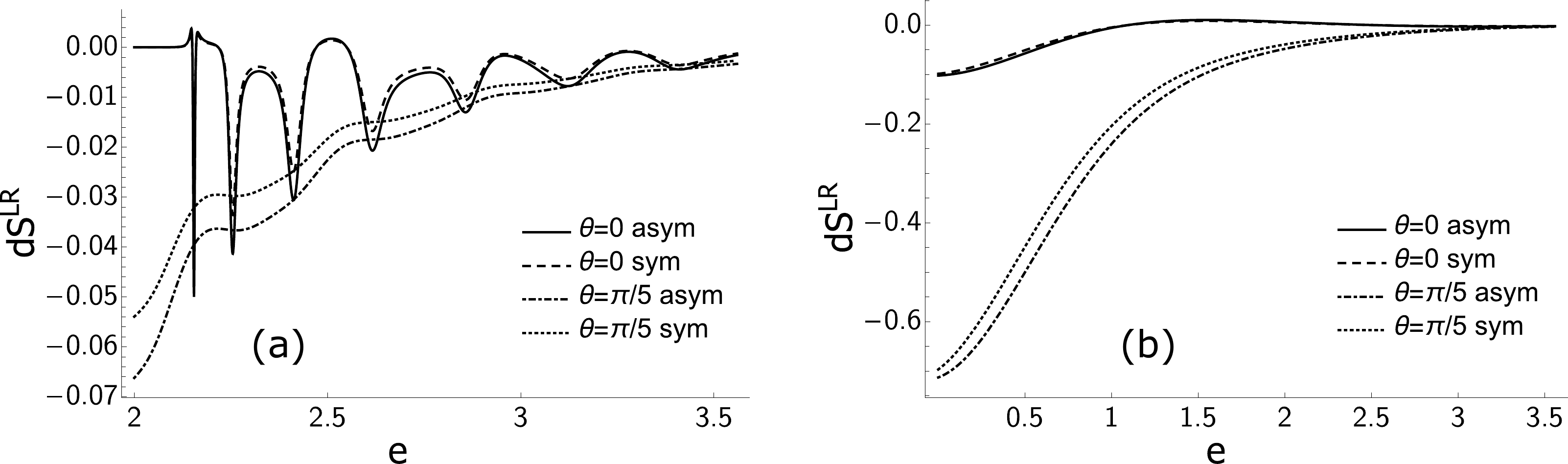}
	\caption{Plots of the total non-local differential shot noise $dS^{LR}$ as a function of energy and with the cross-sectional slices $\theta=0$ and $\theta=\pi/5$  shown in Figure \ref{fig:nonlocal_noise_angle_resolved_plots}. (a) $L_S=5000~au$ and (b) $L_S=500$. For each case we show the non-local shot noise for both symmetric and asymmetric biasing.  (In atomic units)}
\label{fig:nonlocal_shot_noise_slices}
\end{figure}

\begin{figure}[htbp]
	\includegraphics[width=\textwidth]{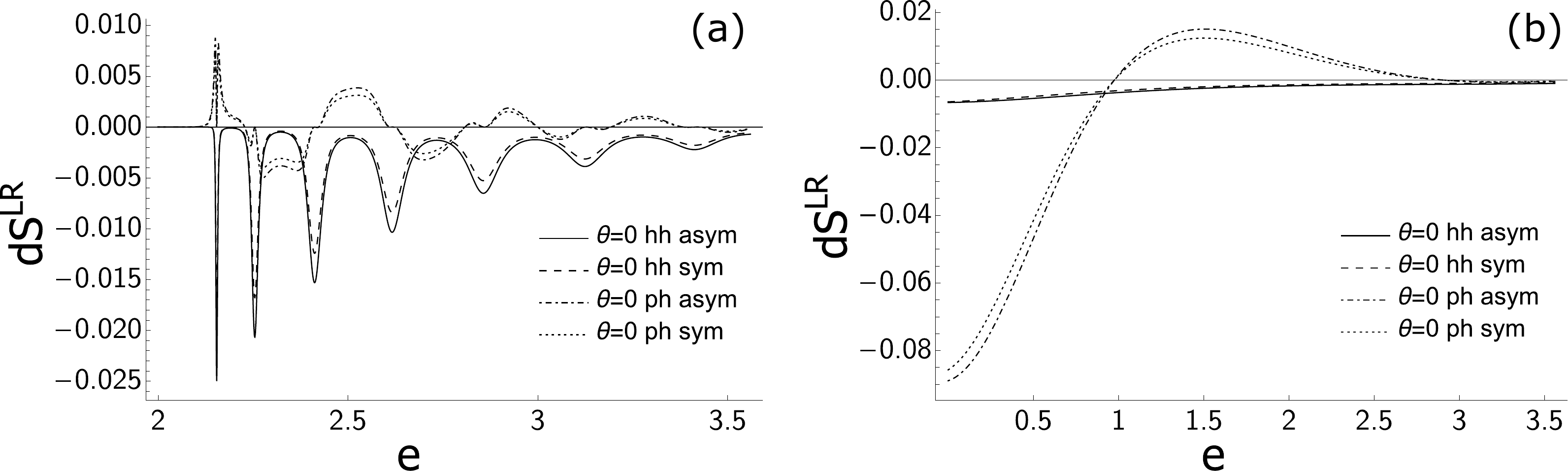}
	\caption{Plots of the nonlocal differential shot noise components $dS_{hh}^{LR}$ and  $dS_{ph}^{LR}$ as a function of energy for  $\theta=0$ and for both symmetric and asymmetric biasing. (a) $L_S=5000~au$. (b) $L_S=500~au$. (In atomic units)}
	\label{fig:nonlocal_shot_noise_slices_comp}
\end{figure}

\begin{figure}[htbp]
	\includegraphics[width=\textwidth]{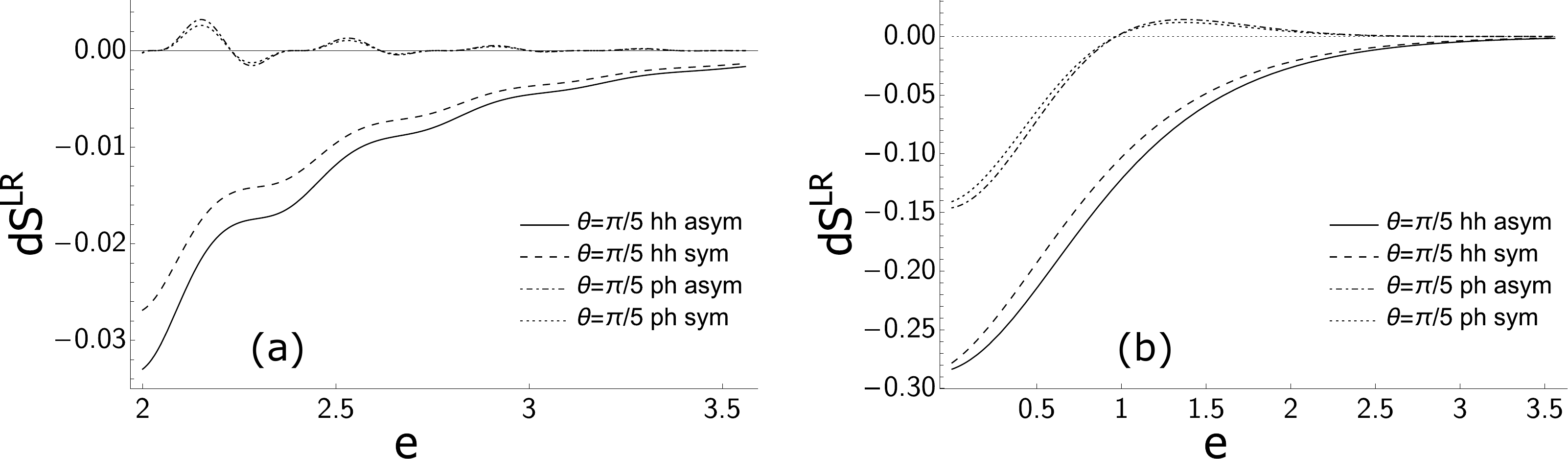}
	\caption{Plots of the nonlocal differential shot noise components $dS_{hh}^{LR}$ and  $dS_{ph}^{LR}$ as a function of energy for  $\theta=\pi/5$ and for both symmetric and asymmetric biasing. (a) $L_S=5000~au$. (b) $L_S=500~au$. (In atomic units)}
	\label{fig:nonlocal_shot_noise_slices_comp_pi5}
\end{figure}

\begin{figure}
	\centering
	\includegraphics[width=.5\textwidth]{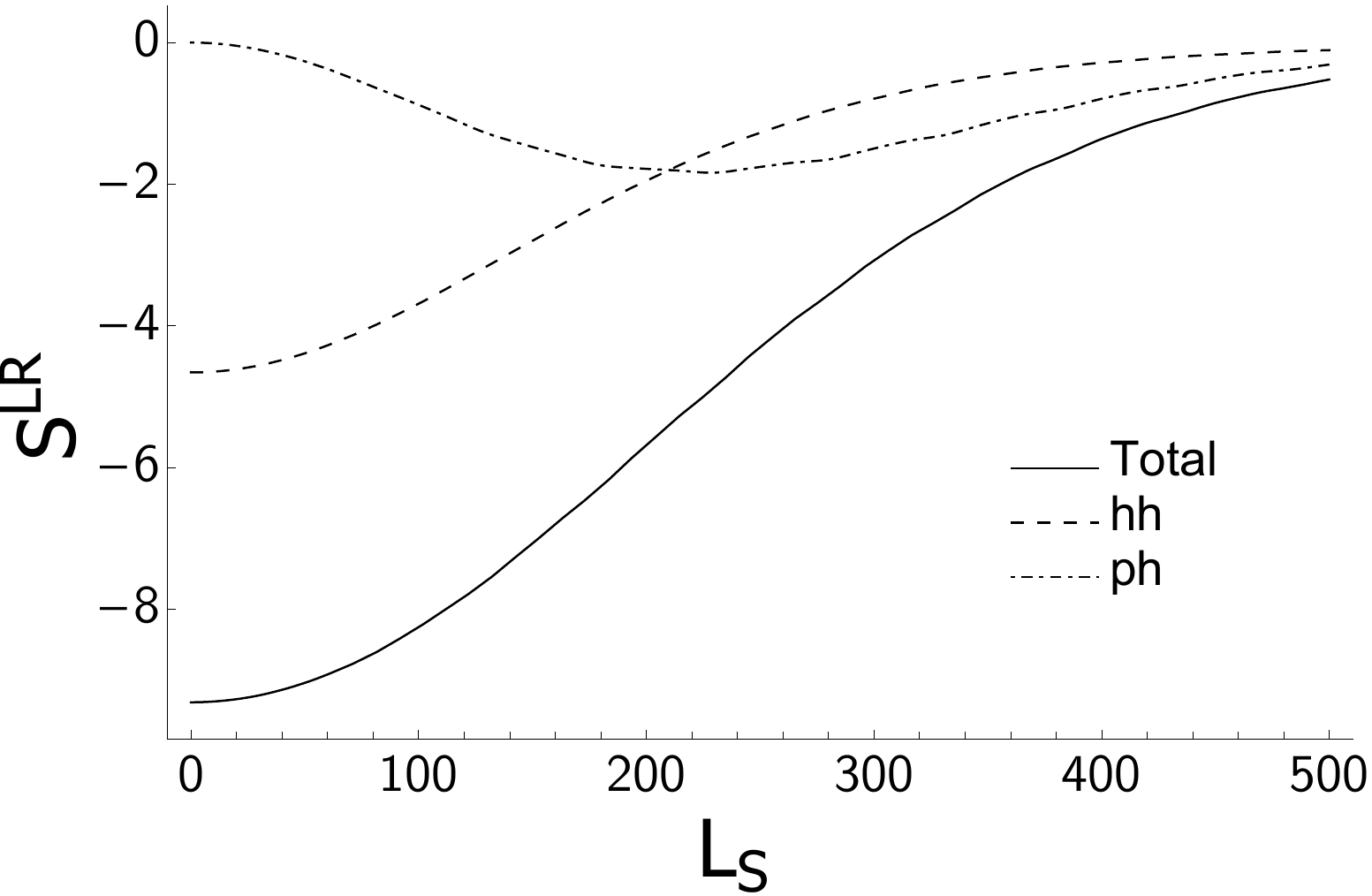}
	\caption{Plot of the total nonlocal shot noise $S^{LR}$  as a function of the length of the superconducting region $L_S$. Values for $S^{LR}$ rescaled by $10^5$ for readability.}
	\label{fig:nonlocal_shot_noise_length_behavior}
\end{figure}

\end{document}